\documentclass[preprint,12pt]{elsarticle}
\usepackage{hyperref}
\hypersetup{
    pdfborder={0 0 2},
    linkbordercolor={0 1 0},
    citebordercolor={0 1 0}
}
\usepackage{fullpage}
\usepackage{amsmath,amsthm,amssymb,graphics}
\usepackage{epsfig}
\usepackage{mathrsfs}
\usepackage{graphicx}
\usepackage{subcaption}
\usepackage{tikz}
\usetikzlibrary{arrows.meta}
\usepackage{color}
\usepackage[margin=0.75in]{geometry}
\usepackage[figuresright]{rotating}
\usepackage{latexsym}
\usepackage{lscape}
\usepackage[rightcaption]{sidecap}
\graphicspath{ {images/} }

\usepackage{array}
\usepackage{enumitem}
\usepackage{float}
\usepackage{placeins}
\usepackage{graphicx}
\usepackage{comment}

\usepackage[percent]{overpic}
\usepackage[utf8]{inputenc}

\large\normalsize

\def\Sa{S}
\def\Ia{I}
\def\Ra{R}
\def\Nb{\mathcal{N}}
\def\Z{\mathbb{Z}}
\def\X{\mathcal{X}}
\def\N{\mathbb N}
\def\I{\mathcal{I}}
\def\P{\mathbb P}
\def\U{\mathrm{U}}
\def\M{\mathcal{M}}

\def\E{\mathbb{E}}
\def\obs{\mathrm{obs}}

\makeatletter
\def\ps@pprintTitle{%
  \let\@oddhead\@empty
  \let\@evenhead\@empty
  \let\@oddfoot\@empty
  \let\@evenfoot\@oddfoot
}
\makeatother

\begin{document}
\begin{frontmatter}

  \title{\textbf{Stochastic network epidemic model and particle filter: \\
General framework and application to influenza in Japan}}

		\author[a]{Ihtisham Ul Haq}
		\ead{ihtisham.ul.haq.d2@math.nagoya-u.ac.jp}
		\author[a,b]{Serge Richard}
		\ead{richard@math.nagoya-u.ac.jp}
	\address[a]{ Graduate School of Mathematics, Nagoya University, Furo-cho, Chikusa-ku, Nagoya, 464-8602, Japan}
    \address[b]{ Institute for Liberal Arts and Sciences, Nagoya University, Furo-cho, Chikusa-ku, Nagoya, 464-8601, Japan}

\begin{abstract}
Parameter inference and state estimation in stochastic and partially observed biological systems 
remain major problems in mathematical biology. 
In this work, we introduce a two-dimensional lattice graph model for the spread of infectious diseases. 
Estimating states and parameters in graph-based stochastic epidemic systems is particularly challenging 
because of randomness and incomplete observations. 
To address these issues, we propose a particle filter based data assimilation framework for the sequential estimation 
of both model states and unknown parameters.
Two methodologies are developed: one based on the number of infected agents and another based on partial spatial location's
information of infected agents on a two-dimensional lattice.
The performance of the two methods are firstly analyzed and validated using synthetic data, 
and the first method is then applied to influenza data 
collected from different prefectures in Japan between July 2024 and December 2025. 
One-week-ahead forecasting simulations are also performed using current weekly data. 
The findings highlight the effectiveness of the proposed PF framework for real-time epidemic monitoring, 
forecasting, and adaptive public health decision-making.
\end{abstract}

\end{frontmatter}

%---------------------------------------------------------------------------------------------------------------------------------------------

\section{Introduction}\label{introduction_section}

Models for describing infectious disease dynamics and assessing the impact of intervention strategies 
can be mainly classified into four main categories: 
stochastic differential equation (SDE) based models \cite{Ikram2022}, 
ordinary differential equation (ODE) based models \cite{AndreuVilarroig2024, AndreuVilarroig2025, Hethcote1989, Kendall1965}, 
partial differential equation (PDE) based models \cite{HaqRichard2025, Novoseltsev2012, Ruan2007, Yang2023}, 
and agent Monte Carlo based models \cite{Hoertel2020, NguyenVanYen2021, Pellis2015}. 
Deterministic models are often used to analyze the speed and pattern of disease propagation. 
In contrast, stochastic models can better capture the variability in real-world data and often provide 
more accurate quantitative and qualitative descriptions of epidemic dynamics \cite{Ball2013, Kadri2025}. 
However, they typically rely on homogeneous mixing assumptions, thereby neglecting 
the underlying physical contact processes between individuals \cite{MendesBaptistaMacNab2024}. 
Theses models are widely used because of their simplicity and low computational cost.

Spatially structured models are widely used to describe real-world problems by local interactions, 
such as epidemic spread, ecological dynamics, cellular automata etc. 
In these models, the system state depends not only on aggregate quantities 
but also on the spatial arrangement of agents \cite{Williams2025}. 
As a matter of fact, during an epidemic any infected individual is more likely to transmit 
the disease to members of their household or to people within their local community \cite{Huber2020}. 
Thus a more realistic approach to epidemic modeling should incorporates complex contact networks, 
where nodes represent individuals and links denote interactions through which infection may spread \cite{IslamUllahKabir2025}. 
Temporal graphs provide another natural framework for representing these time-varying interactions 
and are increasingly used to model realistic human contact networks and epidemic spreading processes \cite{MasudaMillerHolme2021}. 
Since disease transmission occurs through contacts within such networks, 
epidemiological models define explicit transmission rules that govern how infection propagates 
between connected individuals \cite{DimitriouSilvestrosConstantinouPitrisKolios2025}.

As discussed in \cite{Vittadello2022}, parameter inference and state estimation in stochastic and partially 
observed systems remain a major problem in mathematical biology. 
Data assimilation (DA) is a methodology used in science and engineering to estimate hidden states and parameters 
by integrating observational data with models. This approach provides systematic method for combining noisy observations 
and model forecasts within a Bayesian framework, see for example the general DA overview \cite{BachGhil2023}. 
DA has mainly been used for continuous systems. However, its application to discrete-time and discrete-event systems, 
such as agent-based models and graph-based models, remains relatively unexplored \cite{HuangXieChoVerbraeck2023}; 
for  reviews of agent based DA challenges, see 
\cite{GhorbaniGhorbaniNazariHerisAsadi2023, TernesWardHeppenstallKumarKieuMalleson2022}. 

This paper studies the spread of epidemic diseases using a graph-based model defined on a two-dimensional lattice. 
The model is based on a probabilistic cellular automaton type, in which each cell/node of the grid represents 
an individual agent identified by its position $\kappa \equiv (p,q),$ where $p,q \in \Z$. 
Each individual can be in one of three states at time $t$: susceptible $S_t^\kappa$, infected $I_t^\kappa$, 
or recovered $R_t^\kappa$. 
The evolution of the system is based on some unknown parameters that need to be estimated from observation data. 
Data assimilation technics, and more precisely a particle filter (PF) approach
is used for simultaneously running the model and extracting the necessary unknown parameters from the data.
Note that an important feature of the PF approach is that it does not require explicit assumptions on the 
mean or variance of the parameters of the system.

Thus, DA is applied to track unknown states and parameters of the graph based epidemic model. 
To the best of our knowledge, this work is the first one which applies data assimilation to this type of model,
see \cite{SMRT2022} for a related approach on free graphs. 
We propose here two PF algorithms for these estimations. The first algorithm relies only on aggregate observations, 
such as the total number of infected individuals. 
The second algorithm incorporates part of the network topology, or equivalently the least necessary number of agent's generations
before a node can be infected. 

During the simulation process, weights are assigned to all particles, 
with each of them representing an independent simulation. 
After the assimilation of observation data and the estimation of these weights, 
a resampling step is applied to the particles.
In the standard approach of PF, 
highly weighted particles are duplicated while lower weighted particles are removed. 
As a result, this usually decreases the diversity among the particles. 
Also, traditional particle filters use transition probability distributions to generate new particles. 
This makes the diversity of the particle set even more reduced. 
However, when the particles are not diverse enough, they cannot represent the true probability distribution well,
especially when sudden changes take place in the system. 
Our algorithm effectively improves particle diversity based on two specific actions:
1) lower weighted particles are not simply removed but perturbed, 
2) a fitness function is introduced for regulating particle diversity.

Once the algorithms are introduced, a series of experiments using synthetic data are performed, 
and several features of our approach are studied. 
The performance of the estimation methods is evaluated using the mean absolute error (MAE) 
and the mean absolute percentage error (MAPE) with respect to the observed data. 
Subsequently, the model and the algorithms are applied to influenza cases in Japan. 
In this setting, we also perform one week ahead forecasting simulations based on the past weeks and the current week data. 

Let us finally describes the organization of this paper.
Section~\ref{model_section} provides the agent-based model on a two-dimensional lattice. 
Two new PF methodologies are proposed in Section~\ref{PF_methods_section}, 
which is further divided into the subsections~\ref{PF_method1_saection} and~\ref{PF_method2_section}. 
The proposed algorithms are tested using synthetic data to estimate both model parameters and system states, 
and their accuracy is evaluated in Section~\ref{apply_synthetic_data_section}. 
The application of the model and PF methods to influenza cases in Japan is presented in Section~\ref{Influenza_Japan_section}, 
Finally, Section~\ref{conclusion} presents the conclusion of this work.

\section{Network-based model on a 2D lattice}\label{model_section}

We model the distribution of the population using an infinite 2D square lattice to study the spread of an epidemic. 
The infection occurs only between directly connected nodes. 
Note that for large populations, lattice-based models are more precise but computationally more expensive than traditional compartmental ODE models.
For that reason, we constantly try to minimize the computational requirements, and in particular we do not fix
the size of the lattice: it will grow with the size of the epidemics.
Nevertheless, the subjacent grid is going to be $\Z^2$.

Each node $\kappa \in \Z^2$ corresponds to one agent. The edges between the neighboring nodes represent the possible physical contact. 
The position of each agent is indexed by coordinates $\kappa \equiv (p,q)$, with $p,q \in \Z$. 
The state of the agent located at node/position $\kappa$ and at time $t\in\N$ is denoted by $\X_{t}^{\kappa}$. 
Each agent can take three possible states: susceptible, infected, or recovered, defined as
\begin{equation*}
\X_{t}^{\kappa} =
\begin{cases}
\Sa^{\kappa}_{t}, 
& \text{if the agent $\kappa$ is susceptible at time $t$,} \\[0.5mm]
\Ia^{\kappa}_{t,\,\gamma}, 
& \text{if the agent $\kappa$ is infected at time $t$ with infection age $\gamma$}, \\[0.5mm]
\Ra^{\kappa}_{t}, 
& \text{if the agent $\kappa$ has recovered at time $t$,}
\end{cases}
\end{equation*}
where $\gamma\in \{1, 2, \dots, \gamma_{\rm{rec}}-1\}$ corresponds to the number of 
time interval of infection (simply called \emph{infection age}). 
The subsequent state of each agent is determined by the state of the node itself and by the ones of its four adjacent nodes
(von Neumann neighborhood). For that purpose, we define the neighborhood $\Nb^{\kappa}$ of the node 
$\kappa \equiv (p,q)$ as the set of nodes
\begin{equation*}
\Nb^{\kappa} = \big\{ (p-1,q),\, (p+1,q),\, (p,q-1),\, (p,q+1) \big\}.
\end{equation*}

The update of an agent’s state depends on several conditions: 
(i) the current state of the agent, (ii) the number of infected agents connected to it, 
(iii) their respective time spent in the infected state.
For that reason, a susceptible agent $\kappa$ is dynamically created once a first infected node appears in its 
neighborhood $\Nb^{\kappa}$. 
For this susceptible agent, we define 
the number of infected agents in its neighborhood $\Nb^{\kappa}$ at time $t$
as
$$
\I^{\kappa}_{t}= \#\big\{\kappa'\in \Nb^\kappa\mid \X_{t}^{\kappa'}=\Ia^{\kappa'}_{t,\,\gamma} \hbox{ for arbitrary } \gamma\big\}.
$$ 
Then, the transition probability for this agent from time $t$ to time $t+1$ depends on a single time dependent
parameter $\beta_t\geq 0$ and is given by
\begin{equation*}
\P\big(\Ia^{\kappa}_{t+1,\,\gamma} \mid \X_{t}^{\kappa}\big) =
\begin{cases}
\U \leq 1 - \exp\!\left(-\beta_t\frac{\I^{\kappa}_{t}}{4}\right), & \text{if  $\X^{\kappa}_{t} = \Sa_{t}^{\kappa}$,}\\[0.1mm]
1, & \text{if  $\X_{t}^{\kappa}=\Ia^{\kappa}_{t,\,\gamma-1}$ and $\gamma\in \{2, \dots,\gamma_{\rm{rec}}-1\}$,}\\[0.1mm]
0, & \text{if  $\X_{t}^{\kappa}=\Ia^{\kappa}_{t,\,\gamma-1}$ and $\gamma=\gamma_{\rm{rec}}$,}\\[0.1mm]
0, & \text{if  $\X_{t}^{\kappa}=\Ra_{t}^{\kappa}$}.
\end{cases}
\end{equation*}
and 
\begin{equation*}
\P\big(\Ra^{\kappa}_{t+1}\mid \X_{t}^{\kappa}\big) =
\begin{cases}
0, & \text{if  $\X^{\kappa}_{t} = \Sa_{t}^{\kappa}$,}\\[0.1mm]
0, & \text{if  $\X_{t}^{\kappa}=\Ia^{\kappa}_{t,\,\gamma}$ and $\gamma<\gamma_{\rm rec}-1$,}\\[0.1mm]
1, & \text{if  $\X_{t}^{\kappa}=\Ia^{\kappa}_{t,\gamma_{\rm{rec}}-1}$,}\\[0.1mm]
1, & \text{if  $\X_{t}^{\kappa}=\Ra_{t}^{\kappa}$}.
\end{cases}
\end{equation*} 
where $\U \sim \mathrm{Unif}(0,1)$ is a random variable following the uniform probability distribution in $(0,1)$.
Clearly, the parameter $\beta_t$ provides the information about the speed of propagation of the epidemics, 
and therefore corresponds to a transmission rate.
Note also that once a susceptible node turns into infected, then its infection age is set to $\gamma=1$.
The nodes are updated at each unit step according to the above rule.
Therefore, in this network-based framework, transitions from the susceptible state to infectious state occur stochastically.

\begin{figure}[htbp]
\centering
\setlength{\tabcolsep}{3pt}
\renewcommand{\arraystretch}{0}
\begin{tabular}{@{}c c c@{}}
\includegraphics[width=0.42\textwidth,height=0.2\textheight]{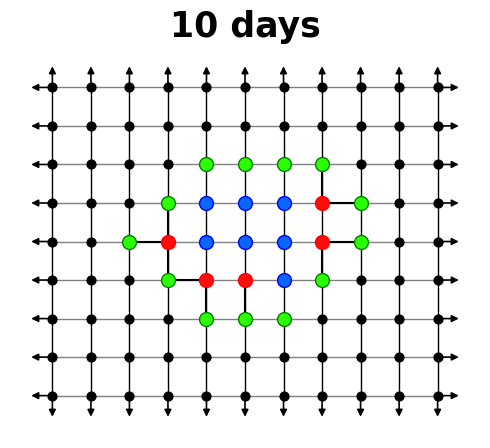} &
\raisebox{5.0\height}{\scalebox{2.5}{$\Rightarrow$}} &
\includegraphics[width=0.42\textwidth,height=0.2\textheight]{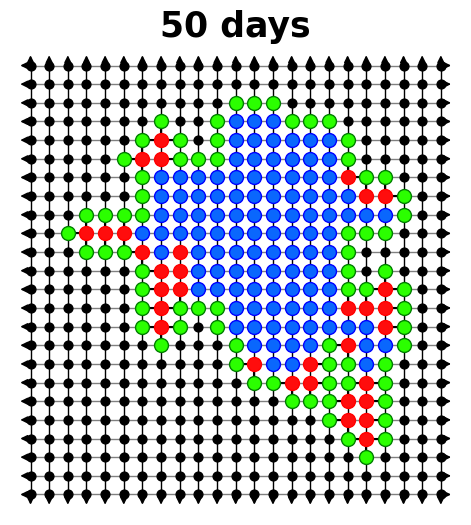}
\\[1pt]
\includegraphics[width=0.42\textwidth,height=0.2\textheight]{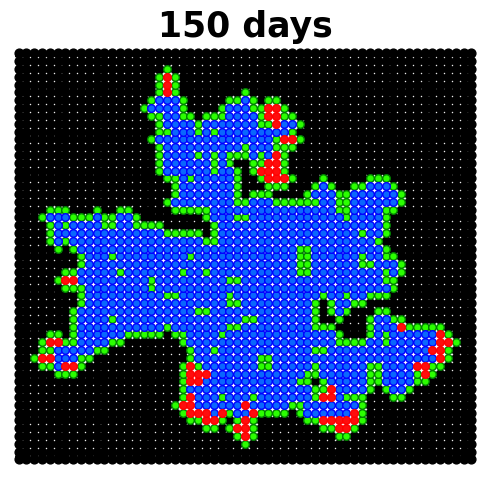} &
\raisebox{4.0\height}{\scalebox{2.5}{$\Leftarrow$}} &
\includegraphics[width=0.42\textwidth,height=0.2\textheight]{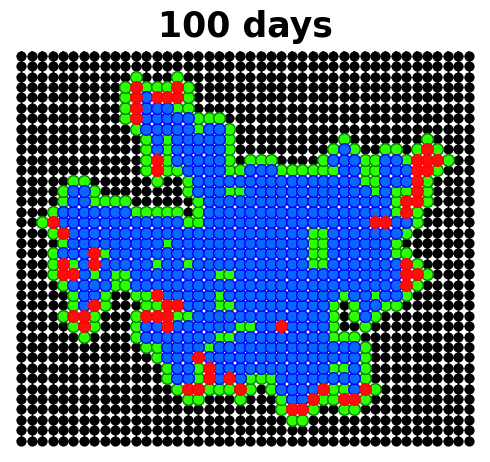}
\end{tabular}
\caption{Spread of infection on a 2D lattice graph-based model with transmission rate $\beta=0.4$ and the recovery time $\gamma_{\rm rec}=8$. 
Susceptible nodes are represented in green, infected nodes are represented in red, while recovered nodes are represented in blue.
Black dots stand for the underlying grid but do not correspond to agents yet.}
\label{model_simulation}
\end{figure}

By gathering the nodes in the different states at time $t$, we can then define three subsets of $\Z^2$
\begin{align}\label{eq_subset}
\begin{split}
\Sa(t) &= \big\{ \kappa  \in \Z^2 \mid  \X^{\kappa}_{t} = \Sa^{\kappa}_{t} \big\}, \\
\Ia(t) &= \big\{ \kappa \in \Z^2 \mid \X^{\kappa}_{t} = \Ia^{\kappa}_{t,\,\gamma} \text{ for some } \gamma \in \{1,2\dots,\gamma_{\rm{rec}}-1\}\big\}, \\
\Ra(t) &= \big\{ \kappa \in \Z^2 \mid \X^{\kappa}_{t} = \Ra^{\kappa}_{t} \big\}.
\end{split}
\end{align}
It then follows that these three sets are always non-intersecting but contiguous, and represent 
all active nodes in our model. The nodes not in these sets do not exist yet, and therefore do not require
any computational time. Also, the boundary of the union of these three sets is always made of susceptible nodes,
are visible in Figure \ref{model_simulation}.
\section{Particle filter methods} \label{PF_methods_section}

Many real-world problems in epidemiology are modeled using complex frameworks, such as agent-based models, stochastic compartmental models, 
or network-based models. For most of them, computing the exact likelihood $p(y \mid x, \theta)$, 
which represents the probability of observing $y$ given the system state $x$ and model parameters $\theta$, 
is extremely difficult. 
In addition, one usually faces the opposite problem, namely deducing the state $x_{t}$ and the model parameter $\theta_{t}$ at time $t$
from the cumulated observations $y_{0:t}$ from time $0$ to time $t$. 
The probability distribution $p(x_{t}, \theta_{t} \mid y_{0:t})$ is referred to as the filtering distribution. 
The particle filter (PF) approach is one method for getting an approximate solution to this problem, as outlined below. 

Let $x_t$ and $\theta_t$ defined as above, and let $y_{t}$ denote the observed data at time $t$. 
The joint posterior distribution of the state and parameter given the observations is then defined as
\begin{equation*}
p(x_{t},\theta_{t} \mid y_{0:t})
\;\propto\;
p(y_{t} \mid x_{t},\theta_{t})\;
p(x_{t},\theta_{t} \mid y_{0:t-1}).
\end{equation*}
We represent the posterior distribution by a set of $N$ particles, each with an associated weight 
$$
\Big\{\big(x^{(i)}_{t}, \theta^{(i)}_{t}, w^{(i)}_{t}\big)\Big\}^{N}_{i=1}, \qquad w^{(i)}_{t}\ge 0, \qquad\sum^{N}_{i=1}w^{(i)}_{t}=1.
$$
Each particle $\bigl(x_t^{(i)}, \theta_t^{(i)}\bigr)$ represents a possible system state, with the associated weight $w_t^{(i)}$ 
reflecting the likelihood of that state given all previous observations. 
The PF approximates this posterior distribution as
\begin{align}
p(x_{t},\theta_{t} \mid y_{0:t})
&\approx \sum_{i=1}^{N} w_{t}^{(i)} \,
\delta\!\left(x_{t} - x_{t}^{(i)}\right)\,
\delta\!\left(\theta_{t} - \theta_{t}^{(i)}\right),
\end{align}
where $\delta(\cdot)$ denotes the Dirac delta function. 
The set of weighted particles
$\{(x_t^{(i)}, \theta_t^{(i)}, w_t^{(i)})\}_{i=1}^{N}$ is obtained by using importance sampling and resampling. 

In the next two subsections we provide more precise explanations about these steps, and in particular
about the resampling, in the framework of the network-based model introduced in Section~\ref{model_section}. 

\subsection{Particle filter: method 1}\label{PF_method1_saection}

We describe here a first approach for getting the posterior distributions. 
Since we focus on the network-based model, the only parameter $\theta_t$ is the coefficient $\beta_t$,
and the state $x_t$ corresponds to the number $\# I(t)$ of infected agents.
Nevertheless, in order to keep some generality in this section, 
we shall use the notations $\theta$ and $x$ for the parameter and for the state.
Note that one additional difficulty not mentioned before comes from the sparseness of the data: 
they are provided only at fixed window size (or window length) $\tau\in \N$ which can be of any length. 
More precisely, the PF estimation is performed at all discrete times determined by
\begin{equation*}
t_k = k \tau, \quad k \in \N=\{0, 1, 2, \dots\}.
\end{equation*} 
Here, $\{t_k\}_{k\in \N}$ are called the filtering time. 
The steps of the first proposed PF algorithm are as follows:

\begin{enumerate}
\item[$(i)$] For $k=0$, an initial ensemble of particles is generated from a prior distribution on the parameter $\theta_{0}$. 
Specifically, for a fixed $N\in \N$ the parameter sample at time $t_{0}$ is constructed over the interval $[a,b]$ as 
\begin{equation}\label{eq_initial_beta}
\theta^{(i)}_0 = a + \frac{b-a}{N-1}\, i,
\qquad i = 0,1,\ldots,N-1.
\end{equation}
For a fixed $m\in \N$, each parameter sample is then duplicated $m$ times, yielding to a total of $N_p = m \times N$ particles, with $m$ an even integer. 
Each particle $i\in \{1, \dots, N_p\}$ is assigned with the same initial condition $x_{0}$ and with a weight $w_0^{(i)}$ equal to $1/N_p$.
Note that the choice of $N$ and $m$ is rather arbitrary and might depend on the computational resources, 
while the choice of the interval $[a,b]$ is based on a vague prior knowledge about the epidemic. 
We finally set $k=1$.

\item[$(ii)$] For $k\geq 1$ and based on the description of the evolution presented in Section \ref{model_section}, each particle
is evolved freely for $\tau$ intervals of time, from $t_{k-1}+1$ to $t_k$. 
During these $\tau$ steps, neither the parameter $\theta^{(i)}_t$
nor the weight $w^{(i)}_t$ of each particle is changed. At time $t_k$, an observation $y_{t_k}$ is available.
For this first method, we assume that $y_{t_k}$ corresponds to the number of infected agents.
Then, a distance between each particle and the observation is computed. 
It is obtained by comparing the observed data with the simulated data as
\begin{equation}
d^{(i)}_{t_k} = \bigl| y_{t_k} - x_{t_k}^{(i)} \bigr|, \qquad \quad \forall i \in \{1, \dots, N_{p}\}.
\label{absolute_distance}
\end{equation} 
We immediately select half of the particles, namely $N_p / 2$ particles, according to the smallest distances $d^{(i)}_{t_{k}}$.
After relabeling, the set of selected particles at time $t_k$ is denoted by 
$$
\M_{t_k}=\{1, \dots, N_p/2\}.
$$

\item[$(iii)$] Among the selected particles, we define the median distance $\hat{d}_{t_k}$, 
as well as the absolute deviations 
$$
\bar{d}_{t_k}^{(i)} = \bigl| d_{t_k}^{(i)} - \hat{d}_{t_k} \bigr|, \qquad \quad \forall i\in  \M_{t_k}.
$$
The median absolute deviation is denoted by $\sigma_{t_k}$, and is going to play the role of a  scaling parameter.
Finally. we assign to each particle $i \in \M_{t_k}$ a new weight defined by 
\begin{equation}
w_{t_k}^{(i)} = \frac{1}{c}\;\!\exp\!\left(-\frac{d_{t_k}^{(i)} - d_{t_k}^{\min}}{\sigma_{t_k}}\right),
\end{equation}
where $d_{t_k}^{\min} = \min_{i \in \M_{t_k}} d_{t_k}^{(i)}$ and where the normalization constant $c$ is given by
$$
c=\sum_{i \in \M_{t_k}} \exp\!\left(-\frac{d_{t_k}^{(i)} - d_{t_k}^{\min}}{\sigma_{t_k}}\right).
$$

\item[$(iv)$] Based on the particles selected in $\M_{t_k}$ and on the weight computed above,
the marginal posterior expectations are approximated by
\begin{align}
\hat{\theta}_{t_k}
= \E[\theta_{t_k} \mid y_{t_k}]
&\approx \sum_{i\in \M_{t_k}} w_{t_k}^{(i)} \theta_{t_k}^{(i)}, \\
\hat{x}_{t_k}
= \E[x_{t_k} \mid y_{t_k}]
&\approx \sum_{i\in \M_{t_k}} w_{t_k}^{(i)} x_{t_k}^{(i)}.
\end{align}
 
\item[$(v)$] Resampling is used to replicate or eliminate particles according to their weights, 
yielding to a new set of particles containing the same initial number of particles. 
The particles in $\M_{t_k}$ are divided into two groups of equal size: the group with the lowest distance $d^{(i)}_{t_k}$, 
and the remaining ones with higher distances. The particles in the first group are simply duplicated, while
the particles in the second group are perturbed. More precisely, for each particle in the second group
having a parameter $\theta^{(i)}_{t_k}$, one particle with parameter $\theta^{(i)}_{t_k}+\epsilon$ 
and one particle with parameter $\theta^{(i)}_{t_k}-\epsilon$ are created, where $\epsilon \sim \mathrm{Unif}(-0.15,\,0.15)$.
The original particle is also removed.
Thus, one ends up with a new set of $N_p$ particles, and can update $k$ to $k+1$.

\item[$(vi)$] The process is repeated from step $(ii)$, as long as necessary.  
\end{enumerate}

Clearly, this procedure enables sequential joint inference of parameters and latent states.
In addition, the selection process in step $(ii)$ and the resampling step $(v)$ aim as mitigating particle 
degeneracy by discarding particles with high error and duplicating those with low error. 
Various resampling schemes have been proposed in the literature; see for example \cite{DoucetJohansen2009}. 
In standard particle filters, resampling is performed based on particle weights. 
However, in our approach, resampling is carried out based on estimation errors rather than weights.
This resampling strategy is easy to implement in all situations and is adopted due to its efficiency 
and relatively low computational cost.
In addition, observe that in step $(v)$, the perturbation of particles introduces some freedom which is especially
useful when the system state undergoes sudden changes. 

However, it turns out that for very sudden changes in the system, even more freedom might be necessary. 
In order to deal with this situation, and alternative step $(v)$ has also been implemented, as described below.
The idea is that if too few particles are considered good enough, a larger resampling is applied
to all particles, not only to half of the particles, as mentioned in $(v)$.
The main idea is to evaluate the fitness of each particle of $\M_{t_k}$, defined as
\begin{equation*}
f^{(i)}_{t_k} =\exp\Big(-\frac{|x_{t_k}^{(i)} - y_{t_k}|}{\max\{y_{t_k},1\}}\Big) \ \in (0,1],
\end{equation*}
and to act according to a fixed upper threshold $\alpha \in (0,1]$ for this fitness:

\begin{enumerate}\label{particles_fitness}
\item[($v'$)] If a prescribed percentage $g\in (0,100]$ of the particles in $\M_{t_k}$ are not considered good enough, namely if 
\begin{equation}\frac{1}{|\M_{t_k}|}\sum_{i\in \M_{t_k}}\mathbf{1}\left(f^{(i)}_{t_k} \geq \alpha\right) < \frac{g}{100},
\end{equation}
where $\mathbf{1}$ denotes the indicator function,
then we do not apply $(v)$ but for each $i\in \M_{t_k}$ one creates one particle with parameter $\theta^{(i)}_{t_k}+\epsilon$ 
and one particle with parameter $\theta^{(i)}_{t_k}-\epsilon$ for $\epsilon \sim \mathrm{Unif}(-\delta,\delta)$
for suitable $\delta\geq 0.15$.
The original particle is also removed.
Thus, one ends up again with a new set of $N_p$ particles, and can update $k$ to $k+1$.
\end{enumerate}
 
This approach effectively avoids particle degradation when the system state changes suddenly.
\begin{figure}[htbp]
    \centering
    \includegraphics[width=1.0\textwidth,height=0.7\textheight]{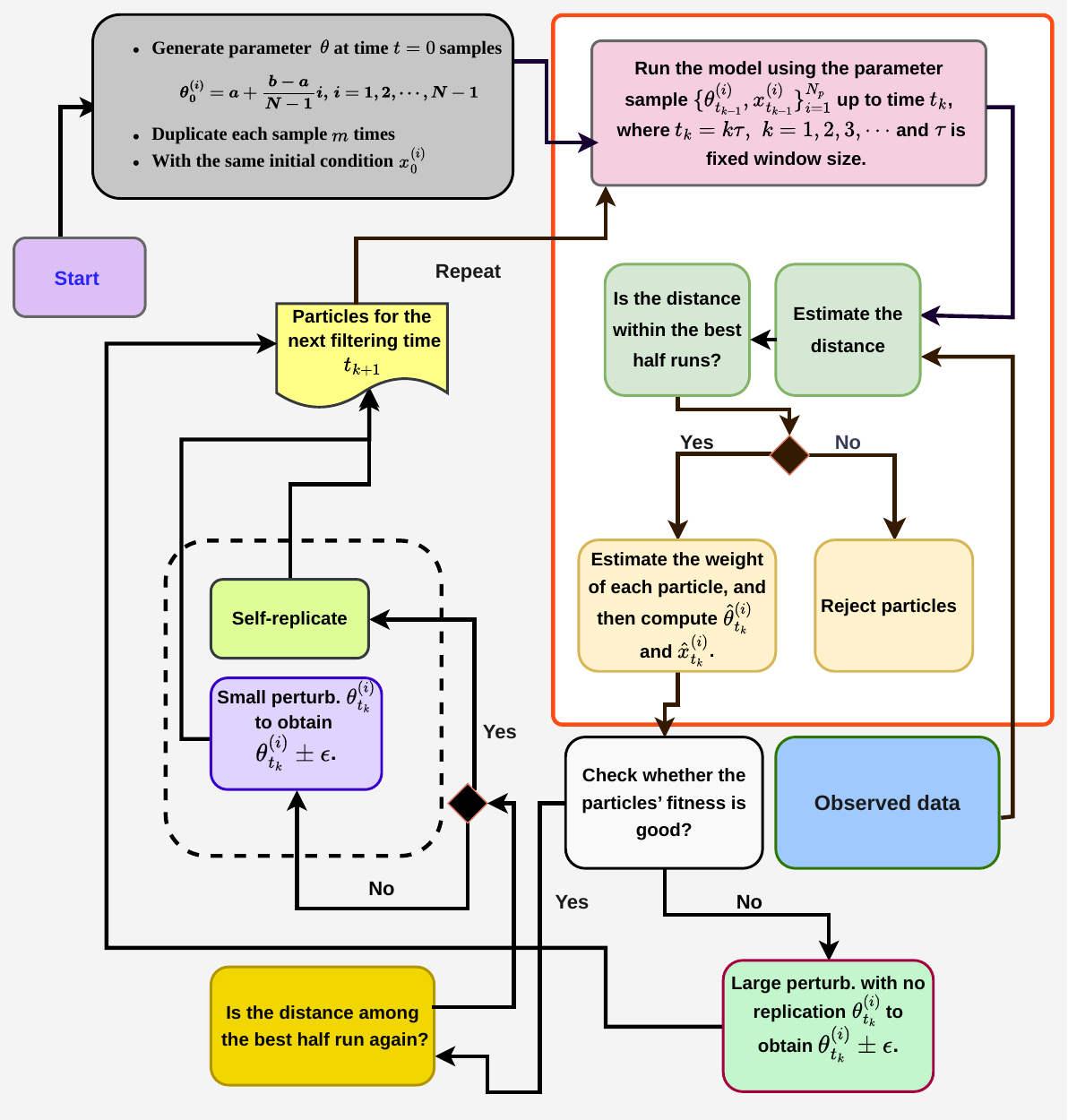}
       \vspace{-0.9cm}
    \caption{Workflow of the PF methods. Both methods begin with same prior sampling and the key difference 
    is how they deal with the observed data and give the weight to the particles. 
    At each time step $t_k$, a new observation $y_{t_k}$ is available.}
    \label{fig:placeholder}
\end{figure}
\subsection{Particle filter: method 2}\label{PF_method2_section}

We propose here a second approach for spatially structured model.
The main part of the algorithm follows the pattern of method 1, but the distance function
is computed very differently. Recall that the position of an agent is denoted by $\kappa=(p,q)\in \Z^2$.
By using the notation introduced in \eqref{eq_subset}, we denote by
$\Ia^\obs(t)$ the positions of all observed infected agents at time $t$. Accordingly, for a particle $i$,
the positions of all infected agents at time $t$ is denoted by $I^{(i)}(t)$.

Let us now partition these sets according to the Manhattan distance (or $\ell^1$-distance) with respect to the origin.
More precisely, for any set $I(t)$ and for any $r\in \N^*$ we define
$$
I_r(t)=\big\{(p,q)\in I(t)\mid |p|+|q|=r\big\}.
$$
Observe that $I_r(t)$ corresponds to part of a diamond shape quadrilateral. 
The interest of using this distance is that it also provides an information about the minimum number of generations of infected agents for 
reaching any agent in $I_r(t)$. More precisely,  for an epidemic starting at $(0,0)$, at least $r$ generations of infected
agents are necessary for reaching any agent belonging to $I_r(t)$.

The second algorithm is based on the same scheme as the first one, except the point $(ii)$, which is replaced
by: 
\begin{enumerate}
\item[$(ii_2)$] For $k\geq 1$ and based on the description of the evolution presented in Section \ref{model_section}, each particle
is evolved freely for $\tau$ intervals of time, from $t_{k-1}+1$ to $t_k$. 
During these $\tau$ steps, neither the parameter $\theta^{(i)}_t$
nor the weight $w^{(i)}_t$ of each particle is changed. At time $t_k$, an observation is available.
For this method, we compare the observed data and the simulated data through the corresponding number of infected
agents at a distance $r$ with respect to the origin, and then sum over $r$.
More precisely we set
$$ 
d_{t_k}^{(i)} = \sum_{r\geq 1} \big| \#I^\obs_r(t_k) - \#I_r^{(i)}(t_k) \big|, \qquad \quad \text{ for } i \in \{1, \dots, N_{p}\}.
$$
We then select half of the particles, namely $N_p / 2$ particles, according to the smallest distances $d^{(i)}_{t_{k}}$.
After relabeling, the set of selected particles at time $t_k$ is denoted by 
$$
\M_{t_k}=\{1, \dots, N_p/2\}.
$$
\end{enumerate}

The other steps of the algorithm correspond to the ones described in subsection~\ref{PF_method1_saection}. 

\section{Experimental setup with synthetic data}\label{apply_synthetic_data_section}

To illustrate the methodology described in the previous sections, we perform a series of experiments using synthetic data. 
Firstly, the model is simulated with known parameters to generate synthetic datasets using the graph-based model 
introduced in section~\ref{model_section}. Then, the two PF algorithms are implemented to estimate the parameter 
and the state of the proposed model, treating the synthetic data as observations.
All algorithms are implemented in Python version~3.13.11.  To evaluate the effectiveness of the proposed approach, we perform the experiments with different filtering window sizes.  In each experiment, the same synthetic dataset is treated as the observed data, and the proposed PF method are applied  to estimate the model parameters and latent states in order to assess the overall performance of the method.

The algorithm employs $N = 20$ parameter particles, each associated with the same initial state, 
and is independently run $m = 20$ times, resulting in a total of $N_{p} = 400$ particles. 
The filtering window size considered is $\tau \in \{5, 10, 15, 20\}$ time unit. 
In all experiments, the model parameter $\gamma_{\rm rec} = 10$ remains constant, while a time-varying transmission rate $\beta_{t}$ is introduced. 
For this experiment and for method 1, we fix the upper threshold parameter $\alpha=0.95$ and the additional parameter $\delta=0.3$, since the variation of 
$\beta_t$ are rather slow. In these experiments, we set $g=15\%$, so condition~\eqref{particles_fitness} is triggered when fewer than $15\%$ of the particles are good enough.

\subsection{Experiment 1: Recovering the transmission rate}\label{sec:parameter_estimation}

Our first aim is to estimate the time dependent transmission rate $\beta_t$, by using the two methods 
introduced in Section \ref{PF_methods_section}.
We firstly assume that the true initial transmission parameter $\beta_{0}$ lies in the interval $[0,1]$ at time $t = 0$, 
and fix $N=20$. With \eqref{eq_initial_beta}, we get $20$ values for $\beta_0$, and run independently our simulations
$20$ times.
For all simulations, the initial condition at $t=0$ is given by $1$ infected agent located at $(0,0)\in \Z^2$. 

\begin{figure}[!htbp]
    \centering
    \begin{overpic}[width=0.49\textwidth,height=0.22\textheight]{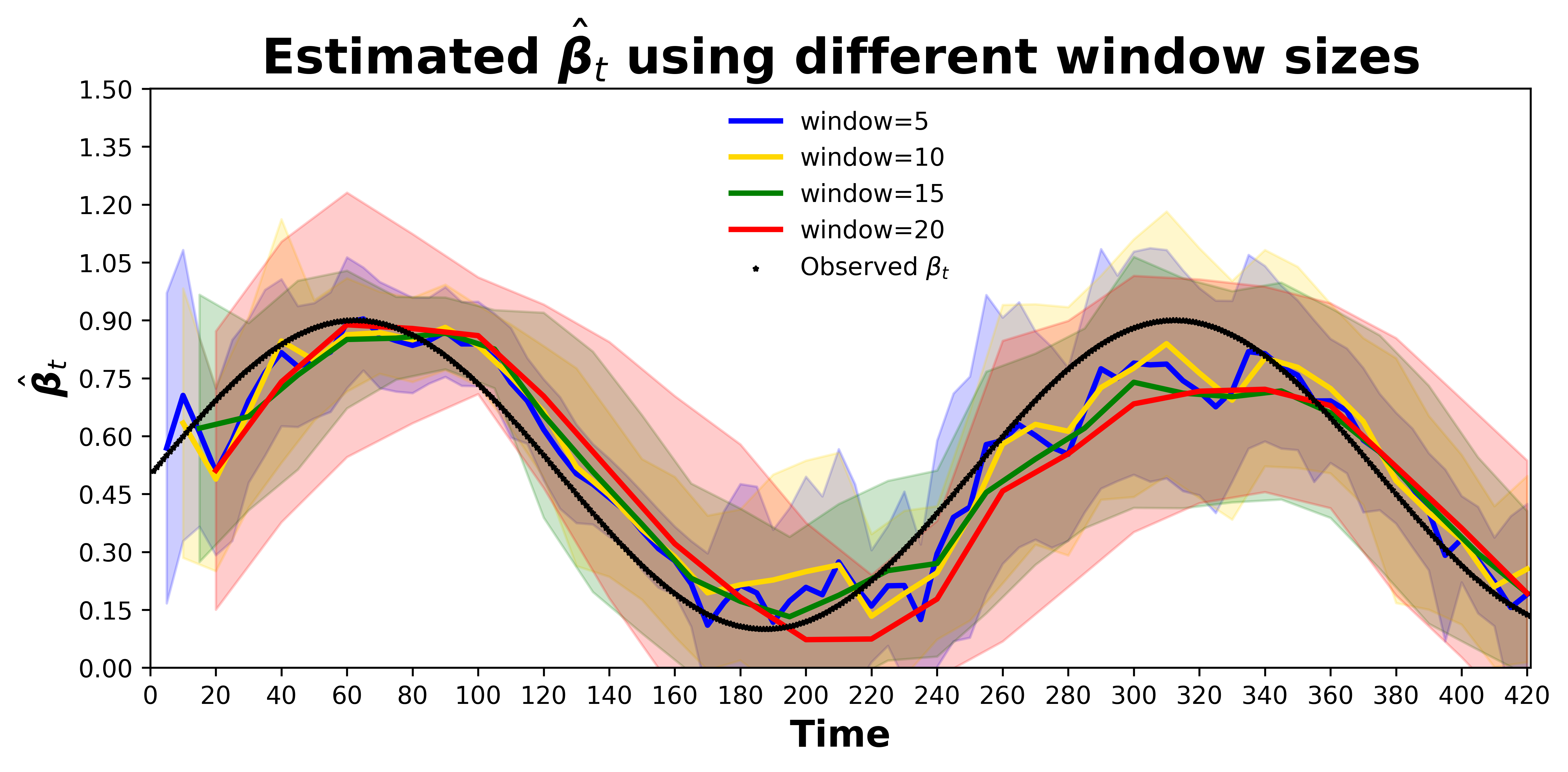}
        \put(75,51){\centering \textbf{\tiny PF method 1}}
    \end{overpic}
    \hfill
    \begin{overpic}[width=0.49\textwidth,height=0.22\textheight]{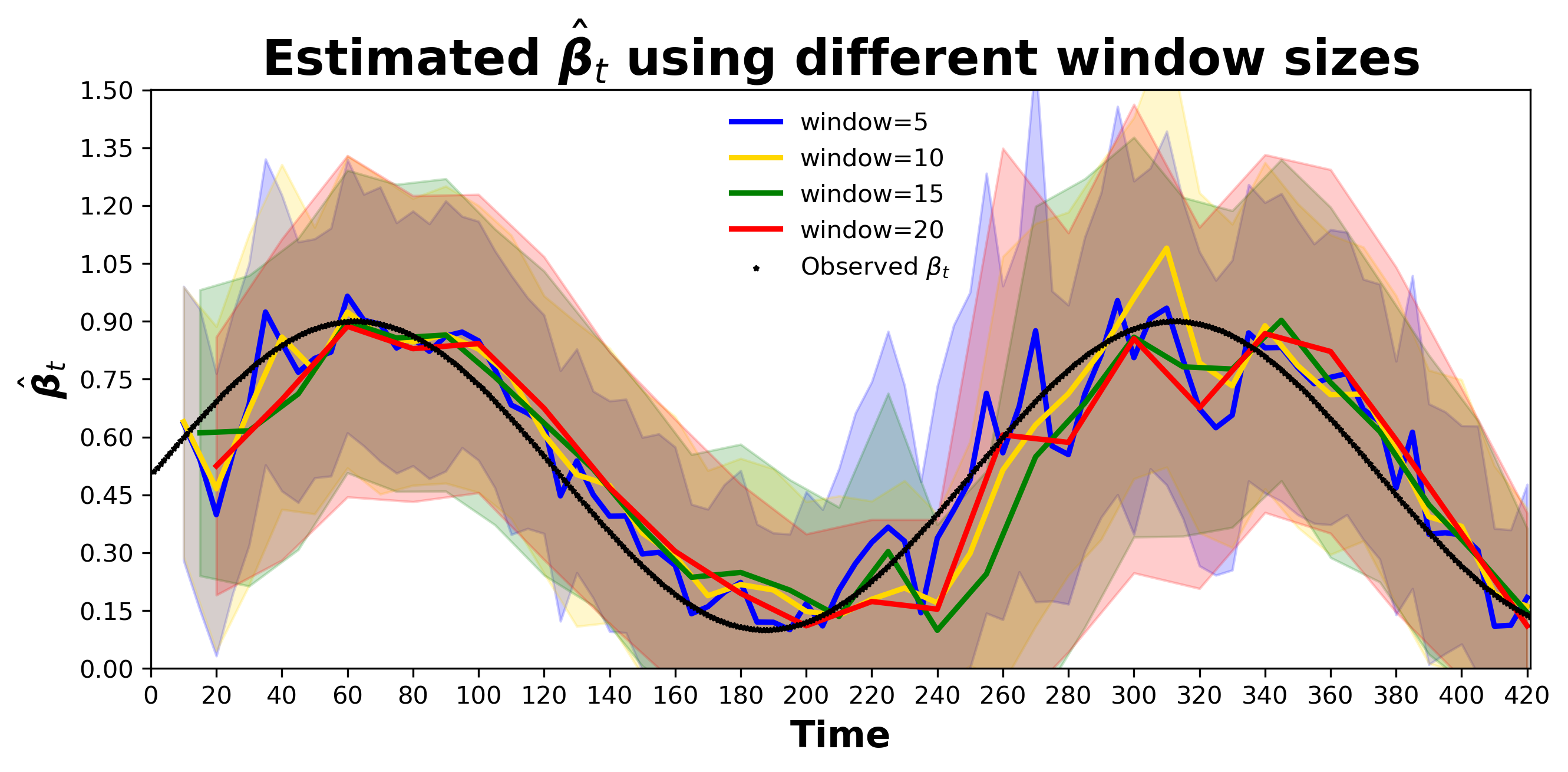}
        \put(75,51){\centering \textbf{\tiny PF method 2}}
    \end{overpic}
    \caption{Behavior of the model parameter estimated using PF methods. The black curve represents the true transmission rate $\beta_t$, 
    while the other colored curves represent the estimated $\hat\beta_{t_{k}}$ values with their $95\%$ confidence intervals 
    obtained using different filter window sizes $\tau$. The recovery rate used in the simulation is fixed at $\gamma_{\rm rec} = 10$. }
    \label{parameter_estimation}
\end{figure}

The estimated results obtained using the PF method 1 and PF method 2 are shown in Figure \ref{parameter_estimation}, 
with a comparison of different filtering window sizes.
The black curve represents the true value of the transmission rate.
The colored curves show the estimates obtained with different filtering windows, 
while the shaded regions indicate the $95\%$ confidence intervals corresponding to each window size. 
The results indicate that the true parameter values fall within the $95\%$ confidence interval, 
closely aligning with the posterior mean. 
However, it can be observed that for window sizes of 10 and less, the performance does not change significantly, 
while the computational cost doubles or even quadruples compared to a window size of $20$. 

We then estimate the Mean Absolute Error (MAE) by using the following formula:
\begin{equation}\label{MAE_formula}
\mathrm{MAE} = \frac{1}{N_{\!f}\,\tau}
\sum_{k=1}^{N_{\!f}} \sum_{j=0}^{\tau-1}
\left|
\beta^{\mathrm{obs}}_{t_k + j}
-
\left(\hat{\beta}_{t_k} + \frac{\hat{\beta}_{t_{k+1}} - \hat{\beta}_{t_k}}{\tau}j
\right)
\right|,
\end{equation}
where $N_{f}$ denotes the number of filtering steps. For this computation we have taken
advantage of the synthetic data by computing the difference for each time $t$ between the true value
and the linear approximation for $\hat{\beta}_t$ shown in Figure \ref{parameter_estimation}.

\begin{table}[!htbp]
\centering
\renewcommand{\arraystretch}{1.1}
\begin{tabular}{ccc}
\hline
\textbf{window} & \textbf{method 1} & \textbf{method 2} \\
\hline
$\tau=5$  & $0.071$ & $0.081$ \\
$\tau=10$ & $0.083$  & $0.079$  \\
$\tau=15$ & $0.089$ & $0.097$ \\
$\tau=20$ & $0.116$ & $0.099$ \\
\hline
\end{tabular}
\caption{Mean Absolute Error for $\hat{\beta}_t$ using \eqref{MAE_formula}.}
\label{MAE_beta_table}
\end{table}

The computed errors across different assimilation windows indicate reasonably accurate estimation for both methods.
Table~\ref{MAE_beta_table} shows that both methods achieve lower errors for window size $\tau=5$, while $\tau=20$ yields higher errors.
Overall, moderate window sizes provide more stable and accurate estimates.

As a comparison, if we compute the mean absolute error only at the observation time,
according to the formula
\begin{equation}\label{MAE_formula2}
\mathrm{MAE} = \frac{1}{N_{f}}
\sum_{k=1}^{N_{f}}
\left|
\beta^{\mathrm{obs}}_{t_k}  
- \hat{\beta}_{t_k} \right|,
\end{equation}
one obtains the rather similar Table \ref{MAE_beta_table2}.

\begin{table}[!htbp]
\centering
\renewcommand{\arraystretch}{1.1}
\begin{tabular}{ccc}
\hline
\textbf{window} & \textbf{method 1} & \textbf{method 2} \\
\hline
$\tau=5$  & $0.073$ & $0.09$ \\
$\tau=10$ & $0.085$  & $0.083$  \\
$\tau=15$ & $0.086$ & $ 0.098$ \\
$\tau=20$ & $0.116$ & $0.101$ \\
\hline
\end{tabular}
\caption{Mean Absolute Error for $\hat{\beta}_t$ using \eqref{MAE_formula2}.}
\label{MAE_beta_table2}
\end{table}

\begin{figure}[htbp]
\centering
\begin{overpic}[width=0.495\textwidth,height=0.2\textheight]{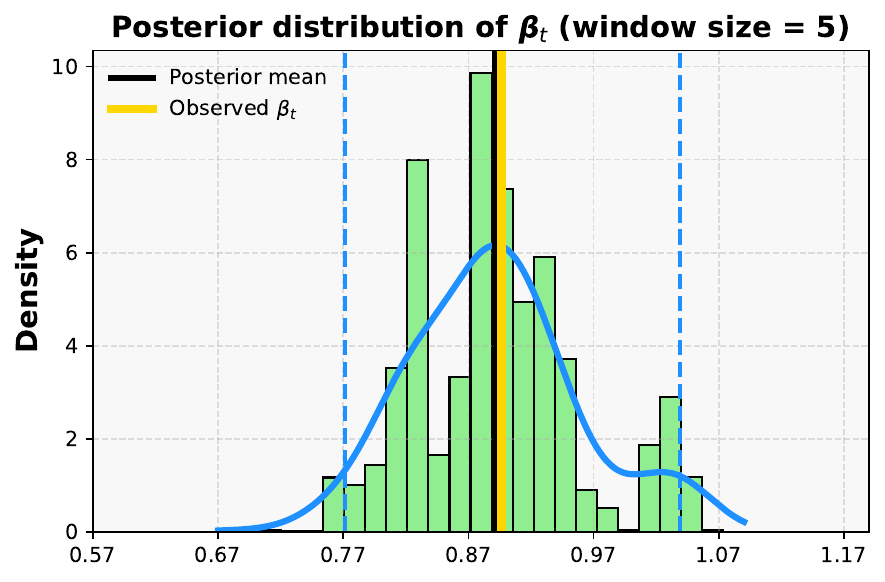}
\put(40,58){\centering \textbf{PF method 1}}
\end{overpic}
\hfill
\begin{overpic}[width=0.49\textwidth,height=0.2\textheight]{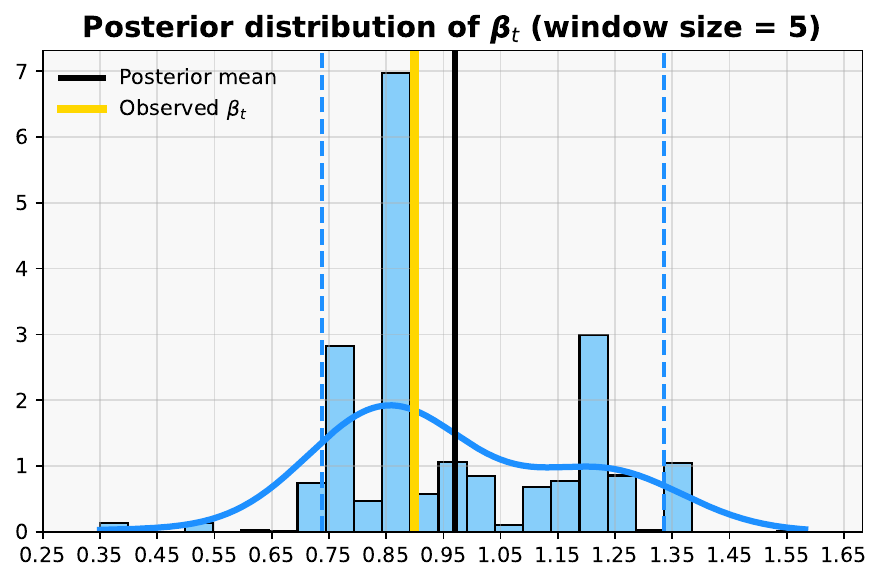}
 \put(40,58){\centering \textbf{PF method 2}}
\end{overpic}
\vspace{0.1cm}
\begin{overpic}[width=0.49\textwidth,height=0.2\textheight]{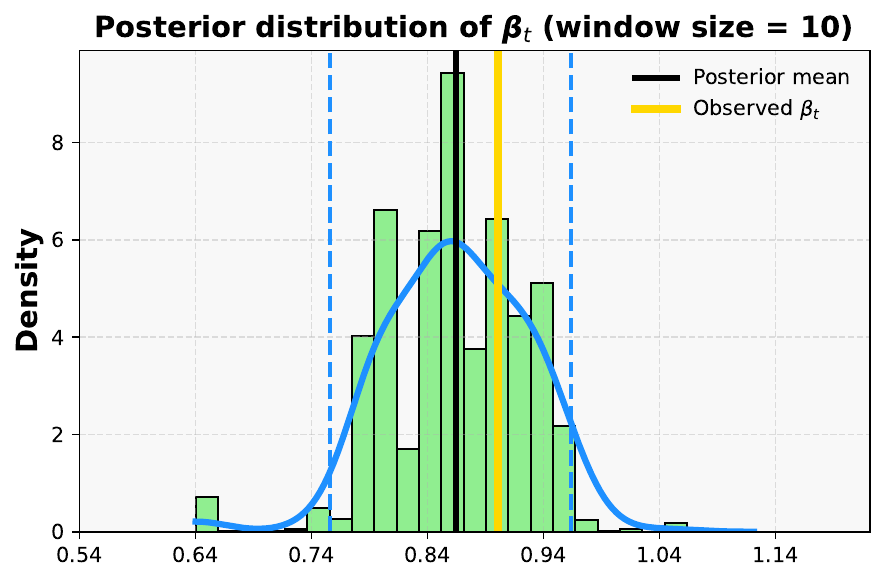}
\end{overpic}
\hfill
\begin{overpic}[width=0.49\textwidth,height=0.2\textheight]{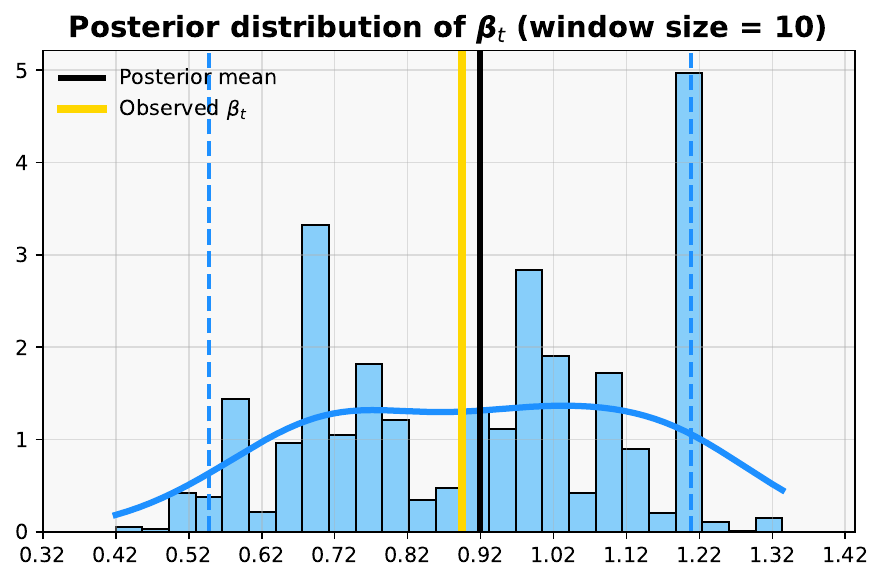}
\end{overpic}
\vspace{0.1cm}
\begin{overpic}[width=0.49\textwidth,height=0.2\textheight]{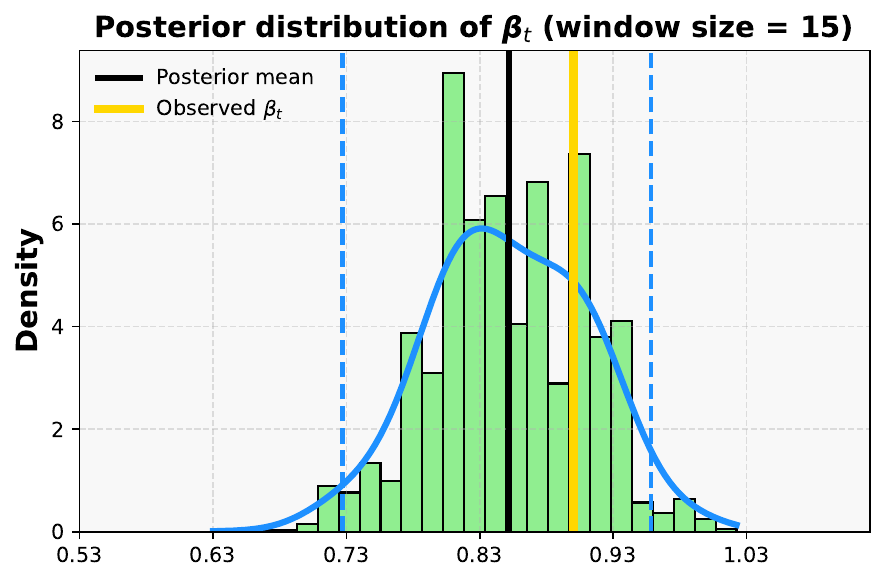}
\end{overpic}
\hfill
\begin{overpic}[width=0.49\textwidth,height=0.2\textheight]{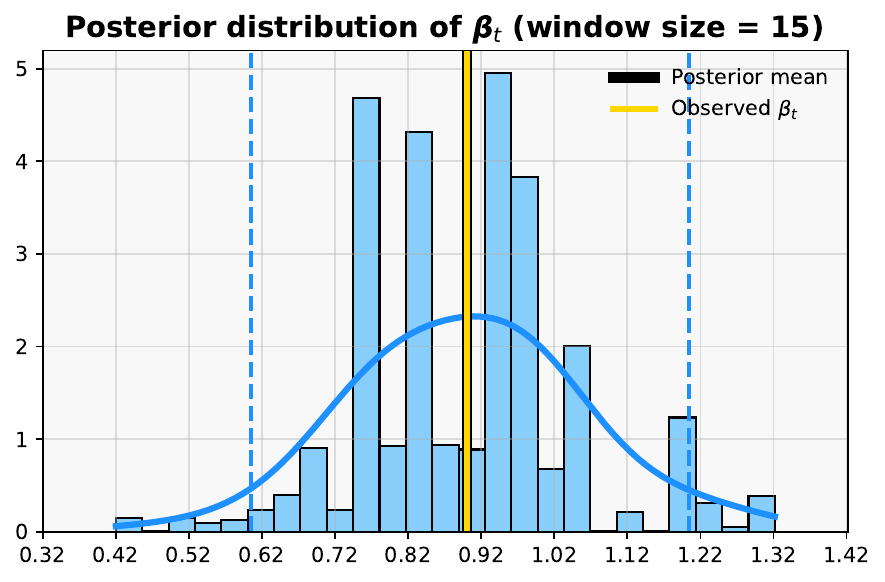}
\end{overpic}
\vspace{0.1cm}
\begin{overpic}[width=0.49\textwidth,height=0.2\textheight]{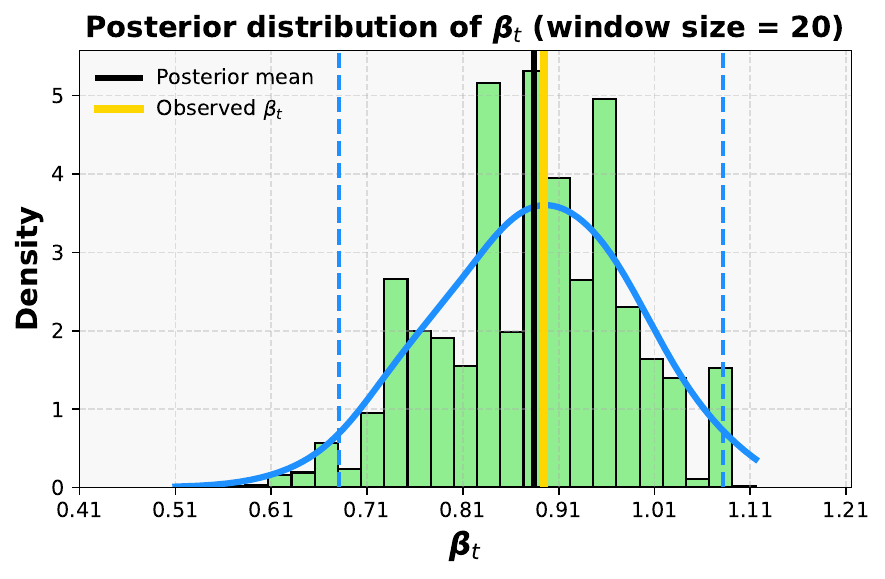}
\end{overpic}
\hfill
\begin{overpic}[width=0.49\textwidth,height=0.2\textheight]{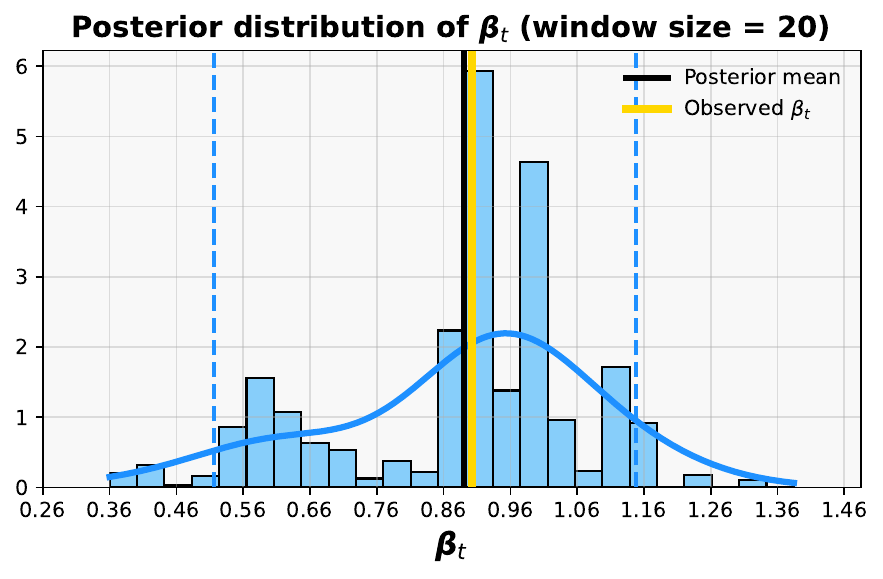}
\end{overpic}
\caption{Posterior distributions of the transmission parameter $\beta_t$ at time step $t_k = 60$.  
The left column corresponds to PF method 1, and the right column corresponds to PF method 2, 
evaluated across different assimilation windows. 
Solid black lines denote the posterior mean estimates, while shaded blue regions represent posterior uncertainty $95\%$ quantiles. 
The blue curves represents the Kernel Density Estimation (KDE) of the posterior distribution of $\beta_{t}$, 
providing a smooth approximation of the distribution. 
Observed parameter values are shown by yellow lines.}
\label{parameter_distributions}
\end{figure}

For the record, we also provide in Figure \ref{parameter_distributions}  
the posterior distribution of the particles at time $t_{k}=60$ for different window sizes.
In these figures, the density of the posterior distribution is represented, 
with the total surface of the rectangles being equal to $1$.
A Kernel Density Estimation (KDE) is also shown, together with the true value $\beta_{60}$,
always lying inside the $95\%$ confidence interval.

\subsection{Experiment 2: State estimation}\label{state_estimation_section}

Our second aim is to estimate the state of the model, by using the two methods 
introduced in Section \ref{PF_methods_section}. Note that for method 1, one can only get
the number of infected agents, while for method 2 information about their spatial distribution
can be inferred.

\begin{figure}[htbp]
    \centering
    \includegraphics[width=0.7\textwidth]{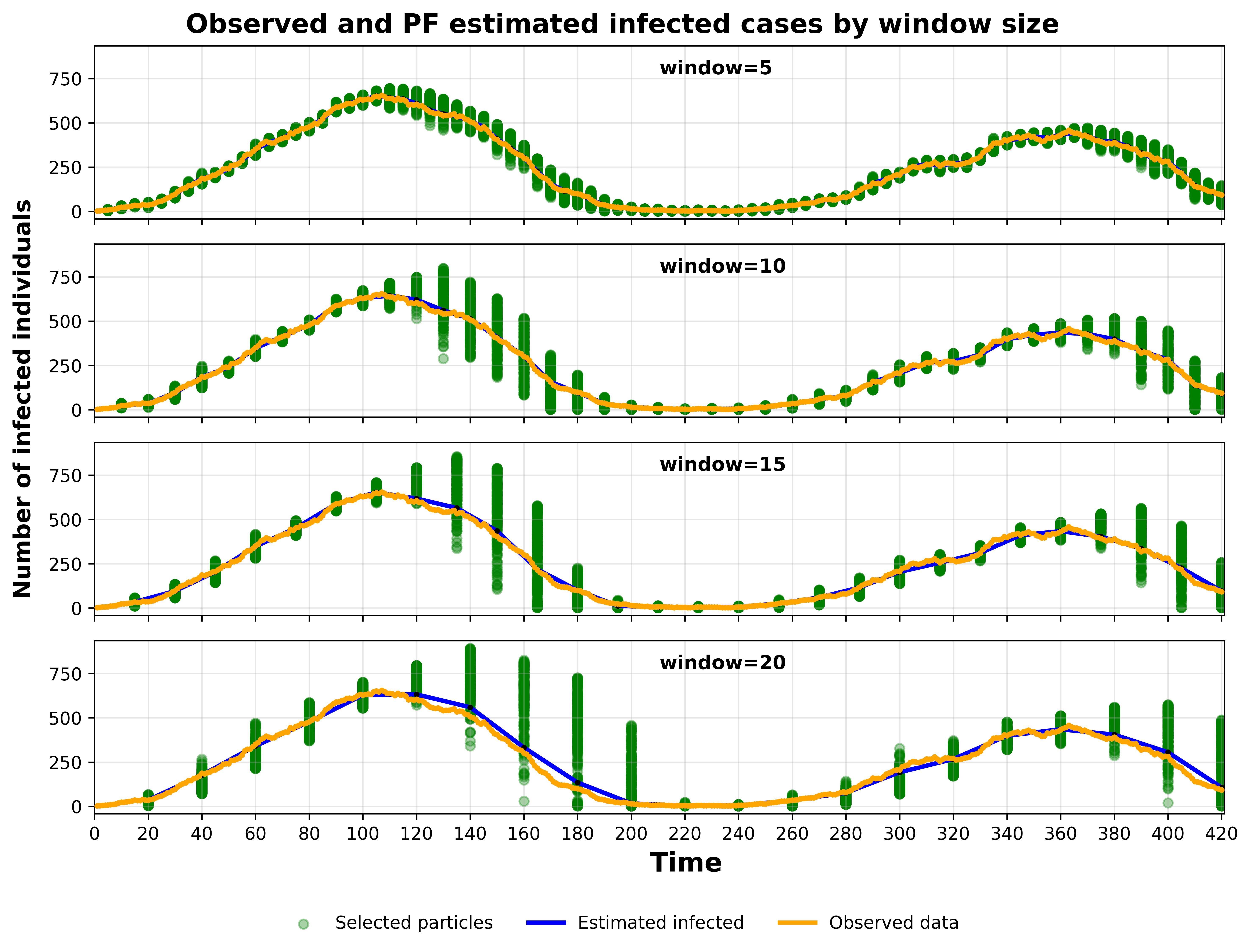}
%    \vspace{-0.7cm}
    \caption{PF method 1 estimation compared with the observed data. The algorithm is applied using window sizes of $5$, $10$, $15$, and $20$.}
    \label{infected_windows_ABL}
\end{figure}

\begin{figure}[htbp]
    \centering
    \includegraphics[width=0.7\textwidth]{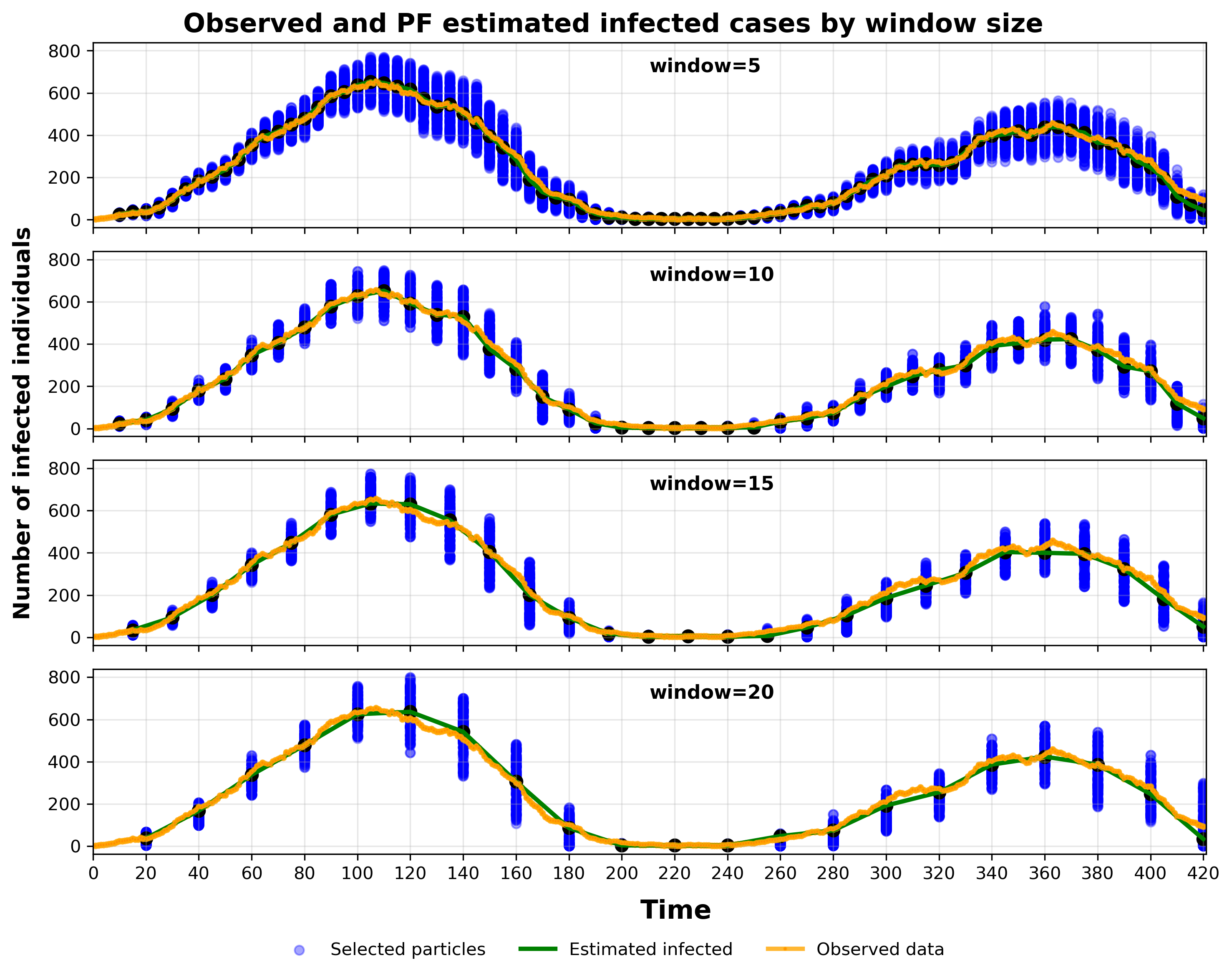}
%       \vspace{-0.7cm}
    \caption{PF method 2 estimation compared with the observed data. The algorithm is applied using window sizes of $5$, $10$, $15$, and $20$.}
    \label{infected_windows_SABL}
\end{figure}

Figures~\ref{infected_windows_ABL} and~\ref{infected_windows_SABL} display the sequential inference of the total number of infected agents
together with the posterior distributions for different window sizes. 
In these figures, the green and blue markers represent $\#I^{(i)}(t_k)$ for the particles selected at the filtering time step $t_k$ 
by PF method 1 and PF method 2, respectively. The corresponding green and blue curves denote the
posterior expectation $\widehat{\#I}(t_k)$ obtained using method~1 and method~2, respectively. 
We can observe that the predicted new cases are in good agreement with the data. 
For a window size of $20$, the limited information available within the interval $[0,20]$ leads to substantial parameter uncertainty. 
In contrast, with a window size of $5$, the inference algorithm captures the system dynamics more effectively, 
resulting in a marked reduction of uncertainty.

\begin{figure}[htbp]
    \centering
    \includegraphics[width=0.8\textwidth]{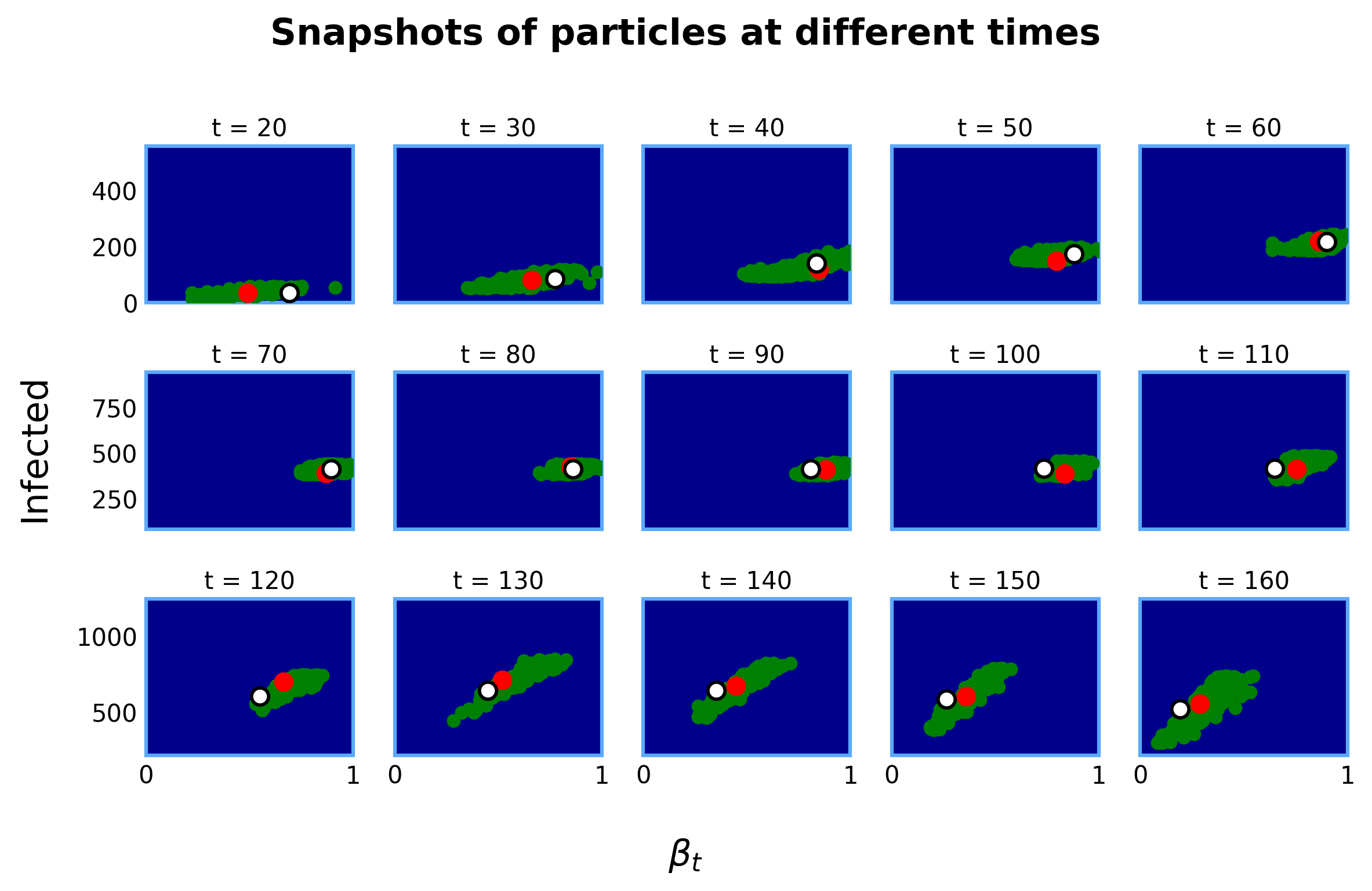}
%       \vspace{-0.3cm}
    \caption{Time snapshots of the distribution of parameters of the selected particles for method 1 for a window size $\tau =10$. 
    Red markers indicate the posterior expectations, and white markers indicate the observations.}
    \label{particles_distribution_ABL}
\end{figure}

\begin{figure}[htbp]
    \centering
    \includegraphics[width=0.8\textwidth]{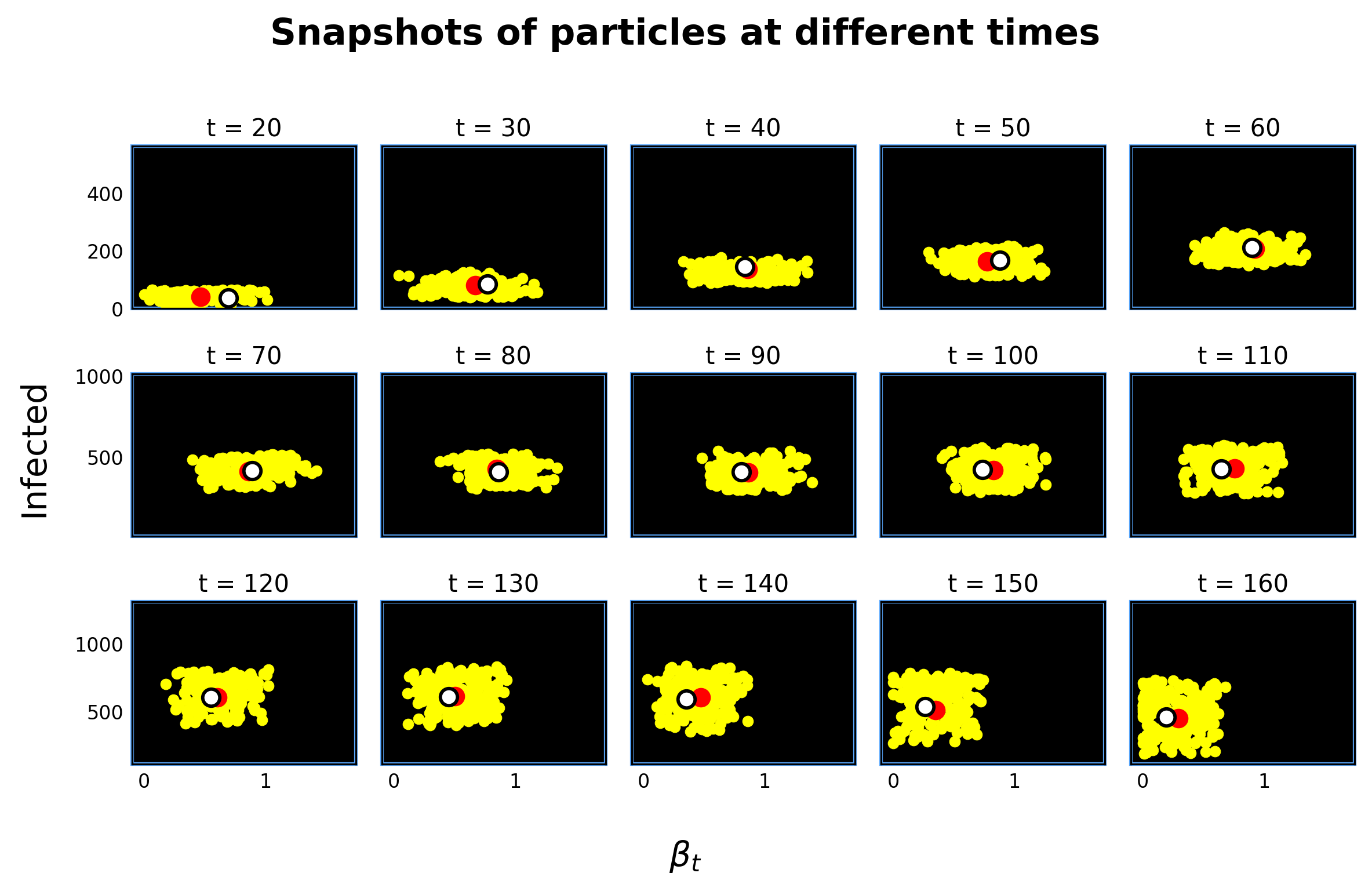}
%       \vspace{-0.3cm}
    \caption{Time snapshots of the distribution of parameters of the selected particles for method 2 for a window size $\tau =10$. 
    Red markers indicate the posterior expectations, and white markers indicate the observations.}
    \label{particles_distribution_SABL}
\end{figure}

Figures~\ref{particles_distribution_ABL} and~\ref{particles_distribution_SABL} illustrate the distribution of particles at selected time steps 
in the PF estimation with a window size of $10$. 
Each panel shows the particles retained by the PF after assimilating the observation at time $t_{k}$. 
The particle samples $\big(\beta^{(i)}_{t_k}, \#I^{(i)}(t_k)\big)$ are shown in green for the PF method 1 case 
and in  yellow for the PF method 2 case. 
The red marker indicates the posterior expected values $\big(\hat{\beta}_{t_k}, \widehat{\#I}(t_k)\big)$, 
while the white marker denotes the corresponding observed data. 
As time evolves, the particles align more closely with the observations, indicating convergence of the filter and a reduction 
of the uncertainty. 
We conclude that the proposed PF method provides accurate joint estimation of model states and parameters.
The particles converge toward the estimates and closely follow the observations, 
indicating reduced uncertainty and stable performance. 
Importantly, good accuracy is maintained even when observations are assimilated only every 20 time steps, 
which lowers the computational cost.

\subsection{Experiment 3: Performance analysis}

\label{PF_performance_section}
Our third aim is to examine the impact of window size $\tau$ on the accuracy of the state estimates. 
Recall that the filter window refers to the number of model time steps after which the filter estimation is performed.
A smaller filter window is expected to provide a better estimate of the true state of the system. 
This is because the model does not need to predict for long periods without support from the observed data.

\begin{figure}[htbp]
    \centering
    \includegraphics[width=0.8\textwidth,height=0.25\textheight]{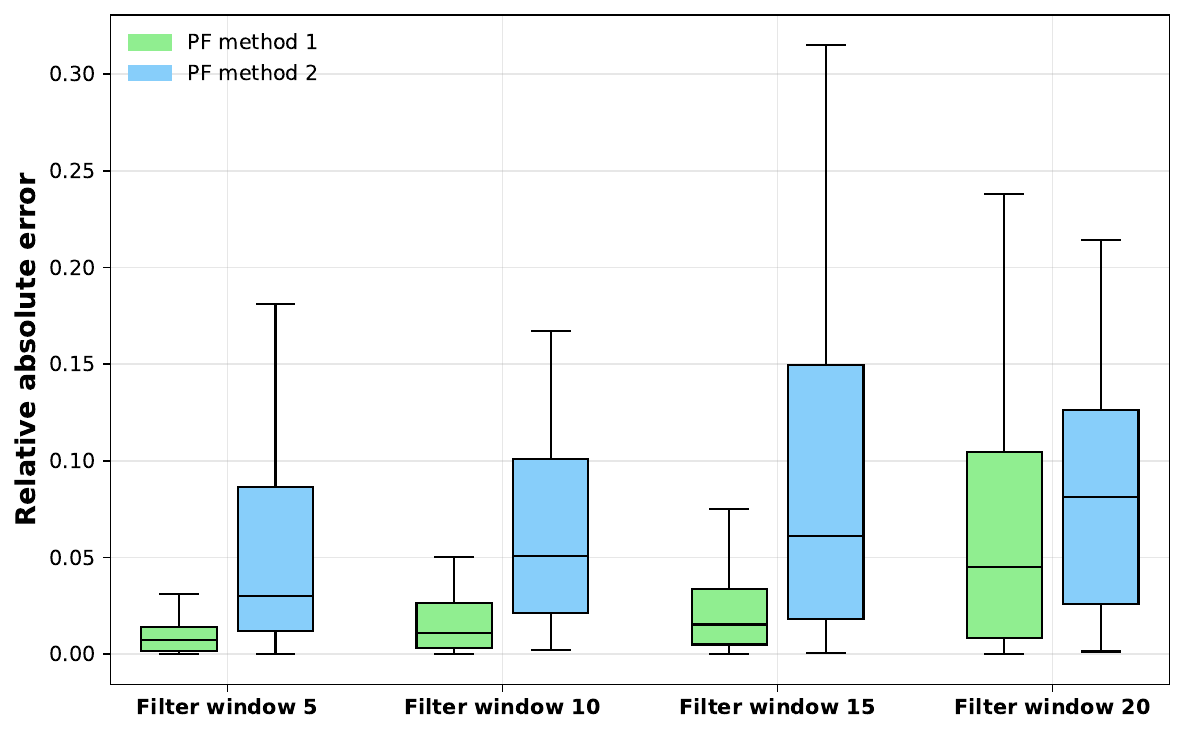}
%       \vspace{-0.3cm}
   \caption{Box plot for the relative absolute error between the estimated forecast and the observed infected agents at all 
   observation times.}
    \label{RMSE_window_veriation}
\end{figure}

We define the \emph{relative absolute error} as the absolute difference between the estimated forecast 
and the corresponding observed value, divided by the observed value.
This quantity is defined for each $t_k=k \tau$ with $k\in \{0,1,2,\dots\}.$
The box plots in Figure  \ref{RMSE_window_veriation} show the distribution of this errors for PF method 1 and 2 
for different filter window sizes $\tau$. Smaller windows reduce the  median errors and variability, 
longer windows increase both the error magnitude and dispersion. 
Clearly, smaller filter windows improve the stability and accuracy of the state estimation.

For information, we provide in 
Table \ref{estimation_performance_table2} the state estimation performance for the number of infected individuals 
for different window sizes $\tau$ and for different time $t_k$.
In addition, we compute the Mean Absolute Percentage Error (MAPE) defined the formula 
\begin{equation}\label{mape}
\mathrm{MAPE}_{\tau} = \frac{1}{N_{\!f}} \sum_{t_k=\tau}^{N_{\!f}}  \left| \frac{\widehat{\#I}(t_k) - \#I^\obs(t_k)}{\#I^\obs(t_k)} \right| \times 100,
\end{equation}
where $N_{\!f}$ denotes the number of filtering steps.
For PF method 1, the MAPE${}_\tau$ increases from $0.46\%$ for $\tau = 5$ to $13.24\%$ for $\tau = 20$. 
This corresponds to an error increase of about $28.8$ times, indicating reduced estimation accuracy as the window size grows. 
Similarly, the PF method 2 shows an increase in MAPE${}_\tau$ from $12.78\%$ to $23.03\%$, 
representing an error increase of about $1.8$ times. 
We observe that PF method 1 consistently achieves lower errors, demonstrating superior 
accuracy in infected-state estimation across all window sizes.
\begin{table}[htbp]
\centering
\renewcommand{\arraystretch}{1.0}
\resizebox{\textwidth}{!}{%
\begin{tabular}{llllll}
\hline
\multicolumn{6}{l}{\textbf{PF method} $\mathbf{1}$} \\
\hline
\textbf{$\mathbf{t_{k}}$} & $\boldsymbol{\tau=5}$ & $\boldsymbol{\tau=10}$ & $\boldsymbol{\tau=15}$ & $\boldsymbol{\tau=20}$ & \textbf{true value} \\
\hline
60  & 350 (313, 386)   & 348 (295, 400)    & 348 (272, 425)       & 342 (188, 495)       & 351 \\
120 & 610 (526, 692)  & 621 (454, 789)     & 618 (374, 862)       & 632 (415, 849)       & 605 \\
180 & 101 (25, 177)    & 97 (2, 216)       & 100 (0, 257)         & 135 (1, 834)       & 100 \\
240 & 7 (2, 12)       & 7 (2, 11)          & 7 (1, 12)            & 5 (1, 10)            & 7 \\
300 & 206 (187, 226)  & 206 (150, 261)     & 204 (126, 283)       & 193 (38, 347)        & 206 \\
360 & 432 (403, 461)  & 435 (378, 492)     & 434 (377, 492)       & 435 (344, 526)       & 432 \\
420 & 92 (34, 150)    & 91 (0, 204)        & 96 (1, 252)          & 107 (0, 520)         &  93 \\
\hline 
\textbf{MAPE}${}_\tau$ & 0.46\% & 1.33\% & 1.09\% & 13.24\% & -- \\
\hline\hline
\multicolumn{5}{l}{\textbf{PF method} $\mathbf{2}$} \\
\hline
60 & 354 (293, 416)        & 344 (256, 432)       & 342 (236, 448)       & 327 (205, 449)         & 351 \\
120 & 616 (484, 748)       & 607 (410, 804)       & 622 (396, 848)       & 644 (370, 917)         & 605 \\
180 & 96 (24, 169)         & 90 (24, 204)         & 82 (2, 199)          & 84 (3, 301)            & 100 \\
240 & 5 (0, 10)            & 3 (1, 6)             & 3 (1, 6)             & 3 (1, 6)               & 7 \\
300 & 209 (147, 272)       & 198 (123, 273)       & 181 (60, 302)        & 186 (37, 366)          & 206 \\
360 & 437 (325, 549)       & 427 (298, 557)       & 409 (268, 550)       & 425 (261, 590)         & 432  \\
420 & 45 (1, 95)          & 48 (0, 122)          & 51 (36, 137)         & 34 (199, 267)          & 93  \\
\hline
\textbf{MAPE}${}_\tau$ & 12.78\% & 17.56\% & 20.45\% & 23.03\% & -- \\
\hline
\end{tabular}}
\caption{Summary of state $\widehat{\# I}(t_k)$: posterior mean, $95\%$ confidence intervals, and true value for different 
window sizes $\tau$ and for the time $t_k \in \{60,120,180,240,300,360, 420\}$.}
\label{estimation_performance_table2}
\end{table} 
\section{Case study: seasonal influenza in Japan}\label{Influenza_Japan_section}

\subsection{Particle filter estimation}

In this section, we use our graph based model approach to describe a real-world case study, 
namely the seasonal influenza in 2024--2025 in some prefectures of Japan. 
In order to estimate the state and the parameter, the PF method 1 is employed. 
For now, method 2 is disregarded, since the notion of epidemic distance or 
generations of infected agents are not available for this study. 

We worked with the seasonal influenza surveillance data provided by the Japan Institute for Health Security (JIHS) 
through Japan’s national sentinel surveillance system for infectious diseases. 
These consist of weekly Influenza incidence cases reported by sentinel doctors at both national and prefectural
levels from July 2024 to December 2025. We performed the simulations for this duration,
with a window size $\tau=7$.
All Influenza data are available in \cite{data}.

For the initial setting, since no prior knowledge was available, the parameter $\beta_0$ is sampled 
in the interval $[0, 7.0]$ following the rule provided in \eqref{eq_initial_beta} with $N=70$.
For each initial condition, we run independently $m=20$ particles, leading to a total of $1400$ particles. 
In all simulations, the recovery parameter $\gamma_{\rm rec} = 7$ is kept constant. 
We set $g=15\%$, so condition~\eqref{particles_fitness} is triggered when fewer than $15\%$ 
of the particles are good enough. Furthermore,  due to the large number of particles, 
we set $\alpha = 0.95$ and $\delta=2$ if the part $(v')$ of the algorithm is required. 

\begin{figure}[htbp]
    \centering
    \includegraphics[width=0.9\textwidth]{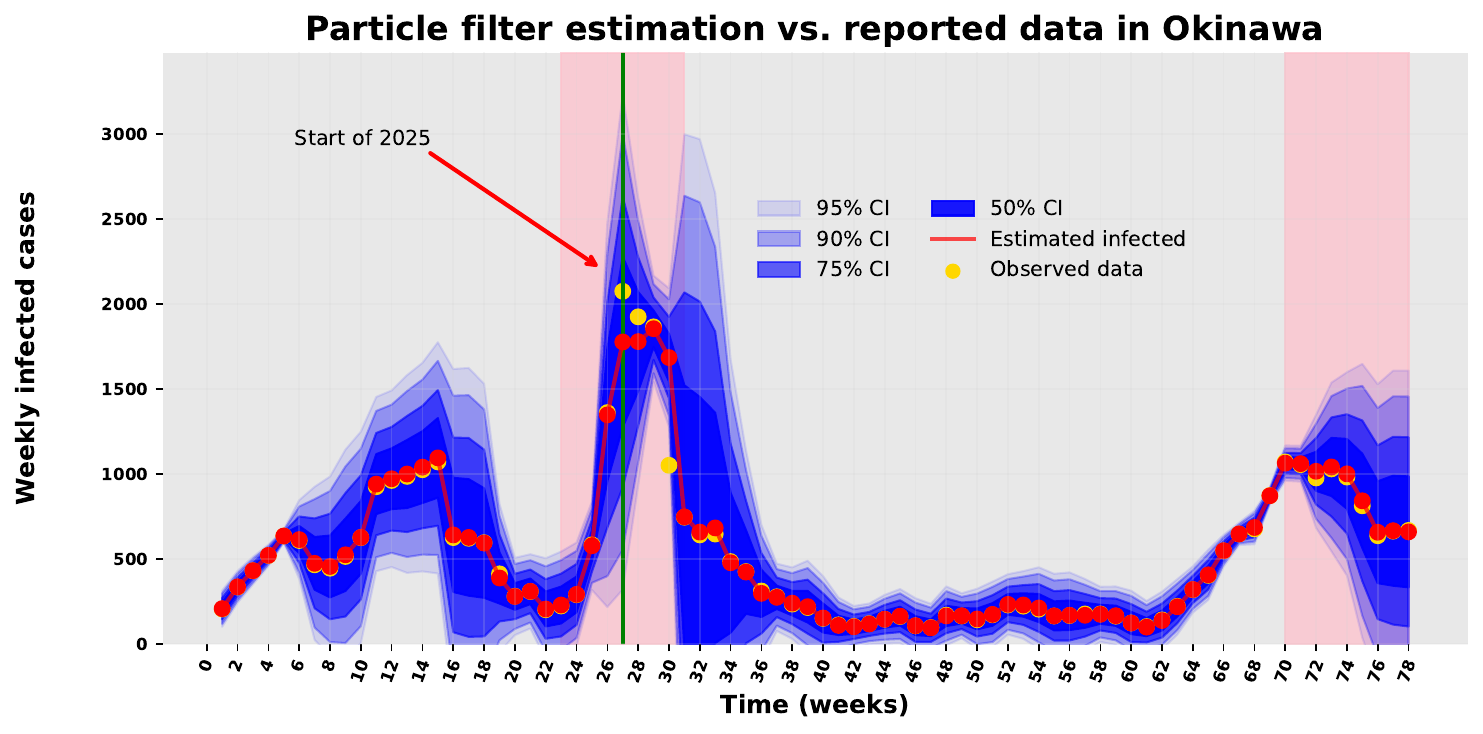}
%       \vspace{-0.3cm}
   \caption{Dynamics of influenza infections in Okinawa estimated using the PF method 1. 
   The estimated red curve is compared with reported cases in yellow dots, 
   and the confidence intervals $50\%-95\%$ highlight the associated uncertainty.}
    \label{Infected_okinawa}
\end{figure}

Figure \ref{Infected_okinawa} contains a comparison of the PF estimated number of newly infected agents 
with the weekly reported influenza data in Okinawa. 
The estimated mean trajectory closely follows the observed data, successfully capturing both the early and late epidemic peaks 
as well as the low-incidence period in-between. 
Almost all observed points lie within the $50\%$ confidence intervals, indicating reliable uncertainty quantification. 
A MAPE value of approximately $2.43 \%$, based on formula \eqref{mape}, indicates excellent estimation performance of the PF. 
Therefore, the PF provides a highly accurate representation of the observed epidemic dynamics.

\begin{figure}[htbp]
    \centering
    \includegraphics[width=0.9\textwidth]{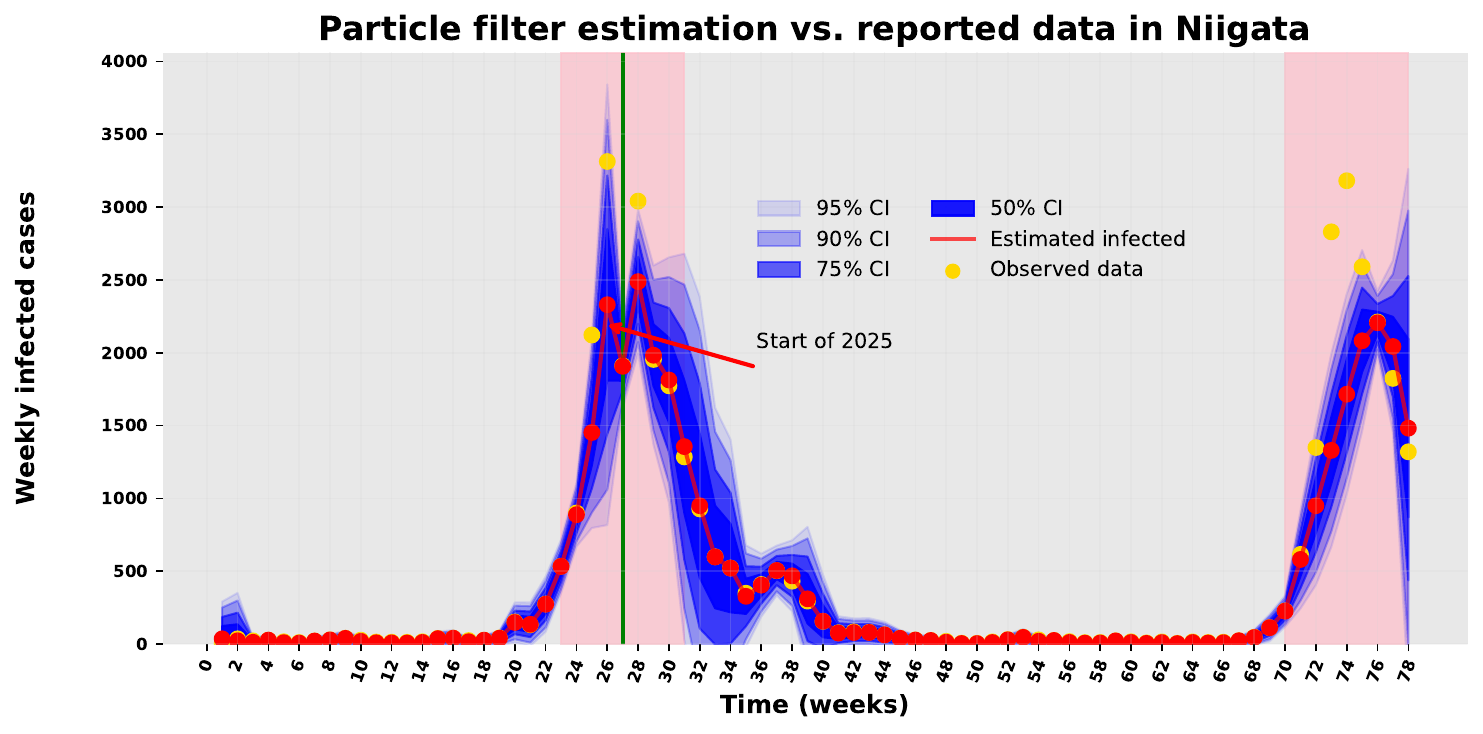}
%       \vspace{-0.3cm}
    \caption{Dynamics of influenza infections in Niigata estimated using the PF method 1. 
    The estimated red curve is compared with reported cases in yellow dots, 
    and the confidence intervals $50\%-95\%$ highlight the associated uncertainty.}
    \label{Infected_niigata}
\end{figure}

In Figure \ref{Infected_niigata}, the PF reproduces the overall seasonal trend for Niigata, including the two main outbreaks.
The observed cases are most of the time covered by the confidence intervals, indicating a consistent uncertainty estimation. 
For comparison, in \cite{SchaumBernalJaquesRamos2022}, the authors used the ensemble Kalman filter for a discrete SIR model 
to forecast COVID-19. Their results show a large delay in the predicted peak over a long period, which we can observe 
in their section 3, Figure 2. The short delay also observed in our results is mainly due to a limitation of the model. 
Infected individuals are assumed to remain static rather than moving spatially. 
Therefore, the infection spreads more slowly in the simulation than in the reported data. 
Especially when the observed cases increase suddenly, the model cannot immediately reach the observed peak. 
Nevertheless, compared with the larger delay reported in \cite{SchaumBernalJaquesRamos2022}, 
our model shows only a short delay of about one to two weeks, despite using weekly data instead of daily data.

\begin{figure}[htbp]
    \centering
    \includegraphics[width=0.9\textwidth]{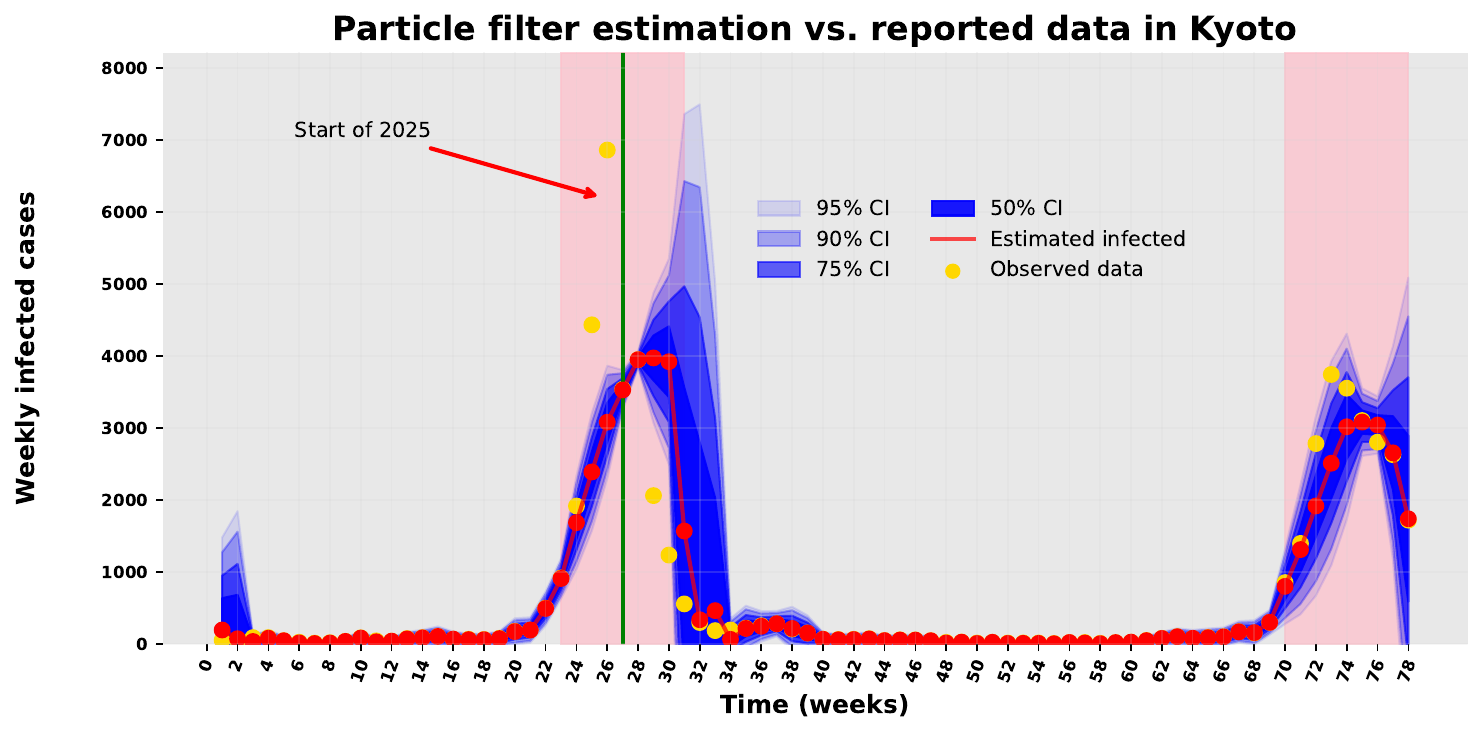}
%       \vspace{-0.3cm}
  \caption{Dynamics of influenza infections in kyoto estimated using the PF method 1. 
    The estimated red curve is compared with reported cases in yellow dots, 
    and the confidence intervals $50\%-95\%$ highlight the associated uncertainty.}
    \label{Infected_kyoto}
\end{figure}

\begin{figure}[htbp]
    \centering
    \includegraphics[width=0.9\textwidth]{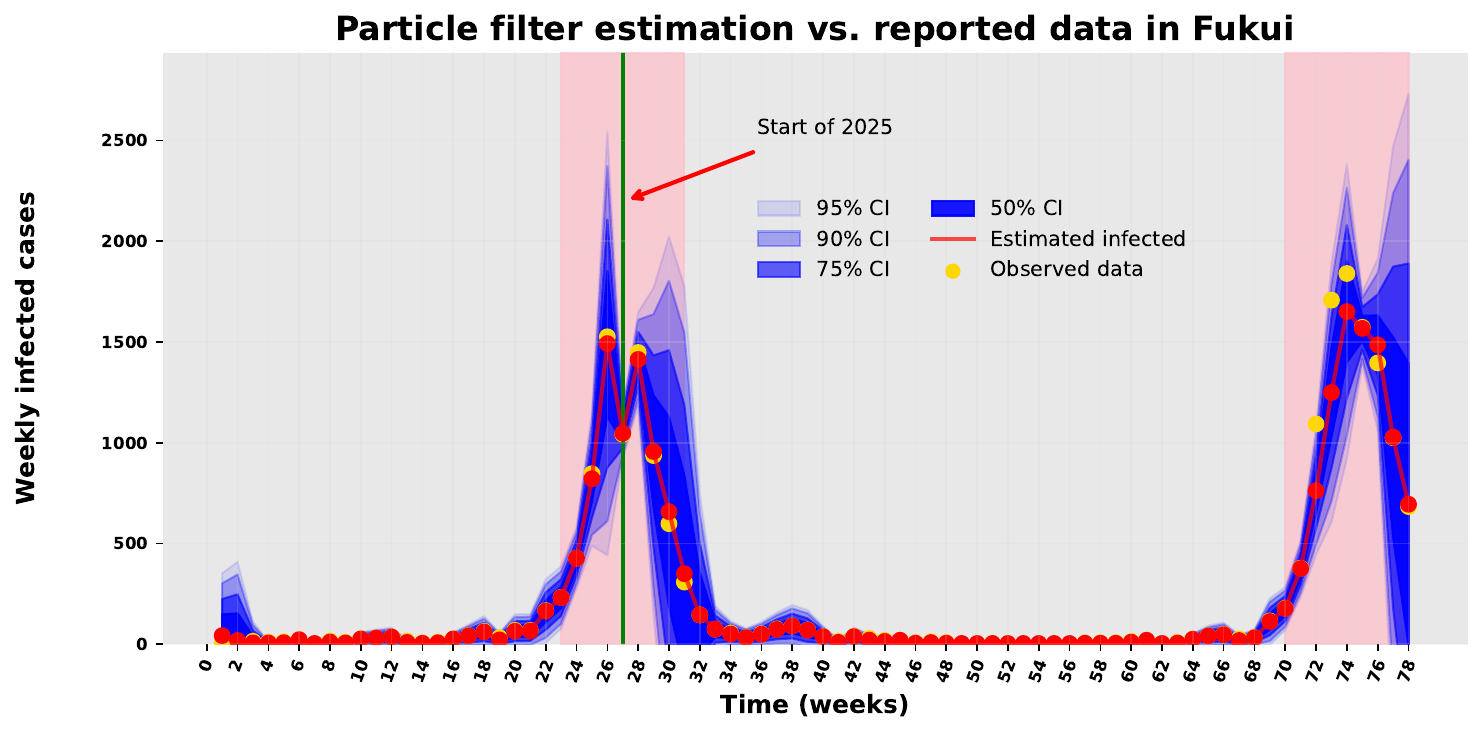}
%       \vspace{-0.3cm}
   \caption{Dynamics of influenza infections in Fukui estimated using the PF method 1. 
    The estimated red curve is compared with reported cases in yellow dots, 
    and the confidence intervals $50\%-95\%$ highlight the associated uncertainty.}
    \label{Infected_fukui}
\end{figure}

Figures~\ref{Infected_kyoto} and \ref{Infected_fukui} show that the PF forecasts 
the seasonal influenza dynamics in Kyoto, and Fukui. 
The model captures the early outbreak, the low transmission period in the middle of the season, and the late epidemic wave. 
In all regions, the estimated mean trajectory closely follows the reported data. 
Most observations lie within the $90\%-95\%$ confidence intervals, indicating a reliable quantification of uncertainties. 

\subsection{One week ahead forecasting}

\begin{figure}[htbp]
    \centering
    \includegraphics[width=0.9\textwidth]{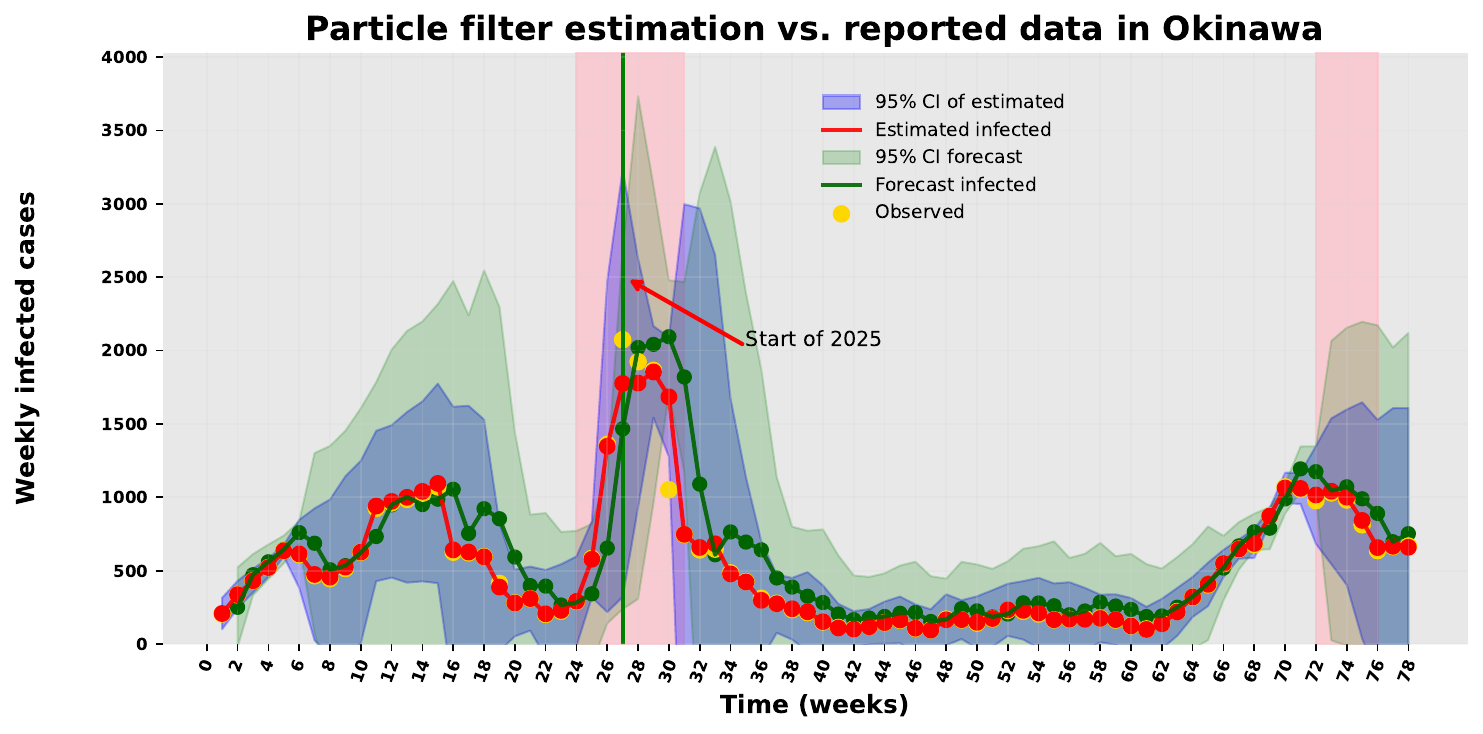}
%       \vspace{-1.0cm}
  \caption{One week ahead model forecast of weekly influenza cases in Okinawa.}
    \label{forcast_okinawa}
\end{figure}

In this section, we analyze the one week ahead forecasts of weekly influenza cases 
and examine the effect of particle filter assimilation on the forecast and posterior distributions. 
In Figure \ref{forcast_okinawa}, the red curve and the blue shaded region represent the particle filter 
estimate and its $95\%$ confidence interval based on the observations assimilated up to the current week. 
The green curve and light green shaded region represent the one-week forecast obtained 
by propagating the selected particles from the previous week forward using the model dynamics and their associated particle weights. 
The yellow dots denote the weekly observed cases. 
This illustrates that the particle filter uses the weighted particle ensemble from the previous assimilation step 
to generate a short-term forecast for the following week. The one week ahead forecast plot indicates that the model 
reproduces the overall temporal pattern of influenza incidence in these four prefectures, 
including the rise, peak, and decline phases of the outbreak. 
However, the forecast intervals become wider near periods of sharper variations, reflecting increased predictive uncertainty. 

\begin{figure}[htbp]
    \centering
    \includegraphics[width=0.9\textwidth]{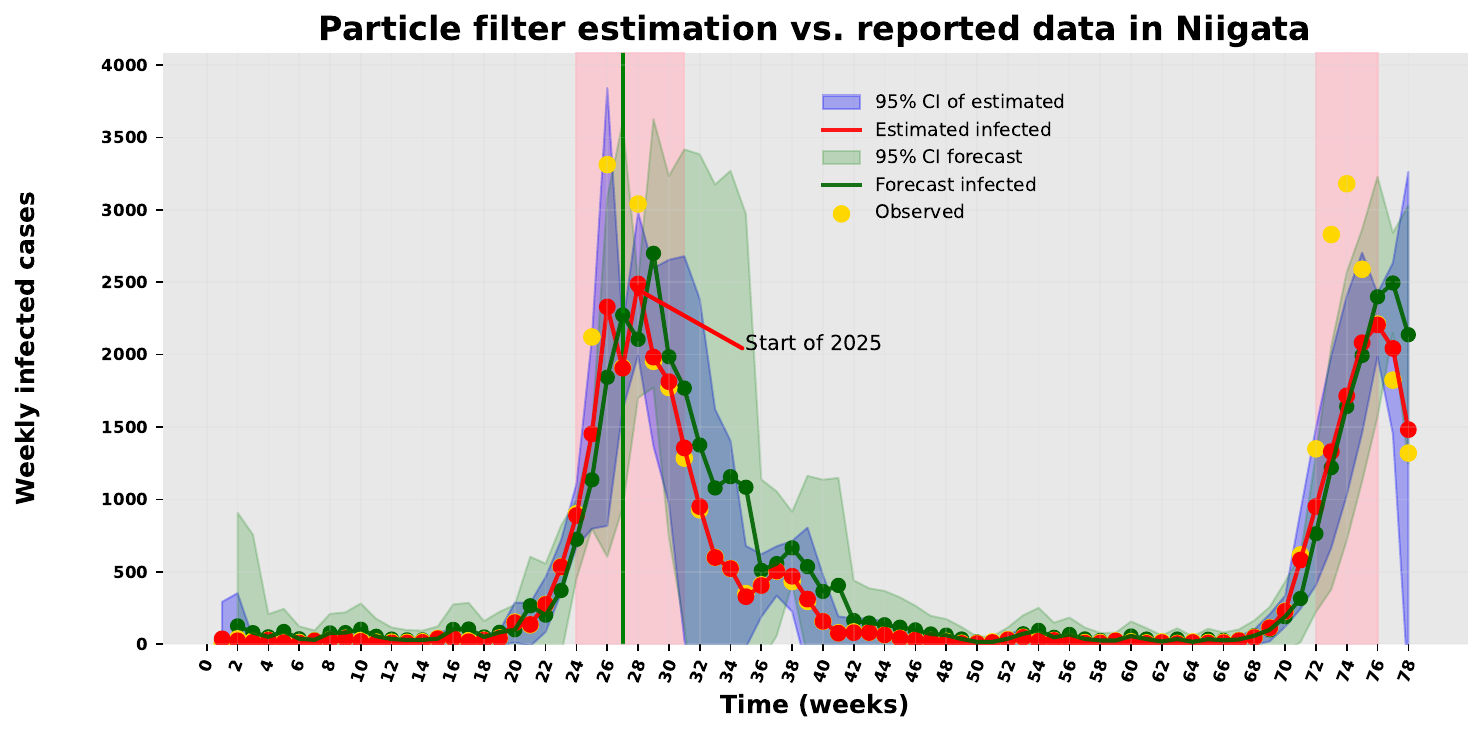}
%       \vspace{-1.0cm}
  \caption{One week ahead model forecast of weekly influenza cases in Niigata.}
    \label{forcast_niigata}
\end{figure} 

The results for Niigata, shown in Figure~\ref{forcast_niigata}, demonstrate the forecasting performance of the model. 
The early outbreak phase, peak time, and decreasing phase are consistently estimated with good accuracy. 
The peak incidence time and the predicted peak incidence are close to the real values. 
The peak incidence is slightly underestimated at week $28$ when projected from week $27$, 
and slightly overestimated at week $29$. 
The predicted peak time in both cases is nearly identical to the real peak time, 
and the general trend is well captured by our model. 
However, at week $74$, we can see that the model is unable to accurately predict the later increase in the incidence curve.

\begin{figure}[htbp]
    \centering
    \includegraphics[width=0.9\textwidth]{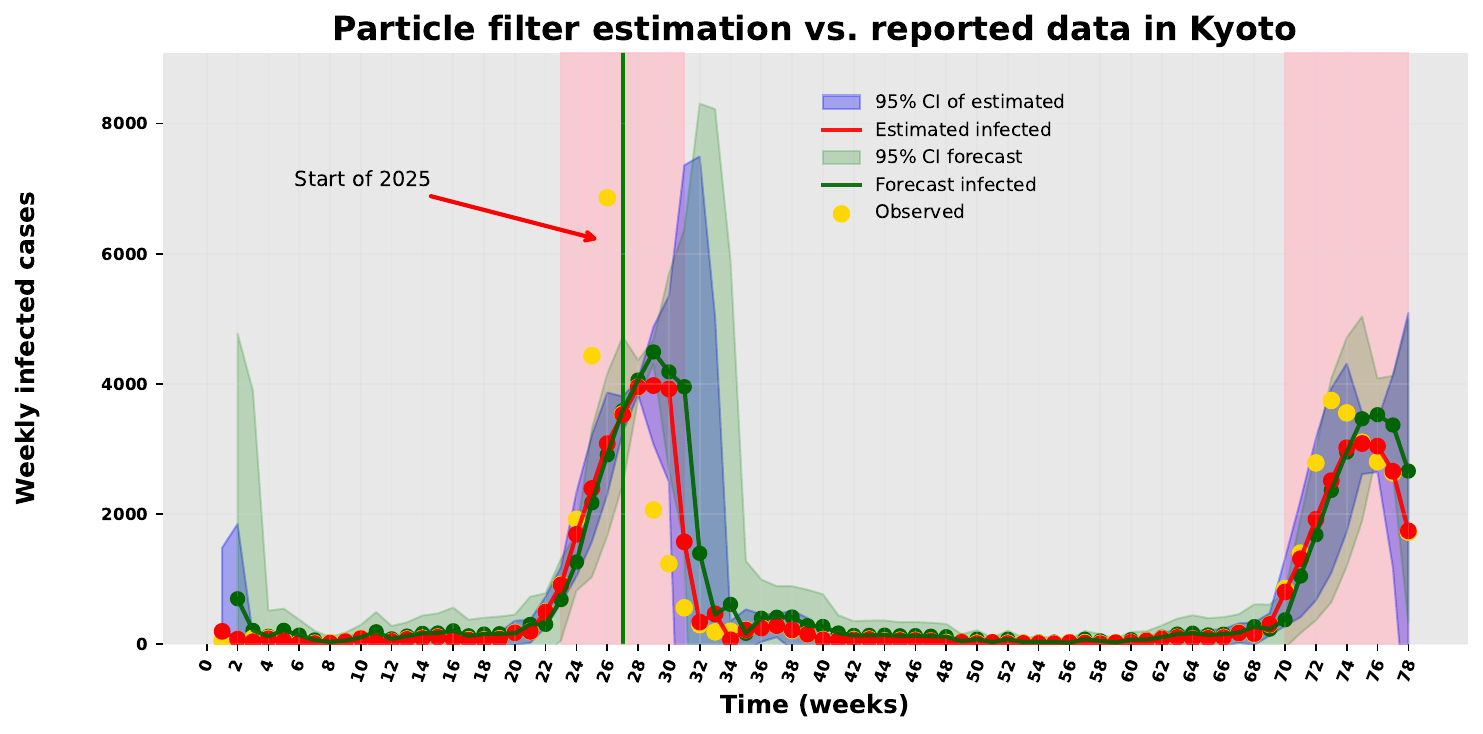}
%       \vspace{-1.0cm}
  \caption{One week ahead model forecast of weekly influenza cases in Kyoto.}
    \label{forcast_kyoto}
\end{figure}

\begin{figure}[htbp]
    \centering
    \includegraphics[width=0.9\textwidth]{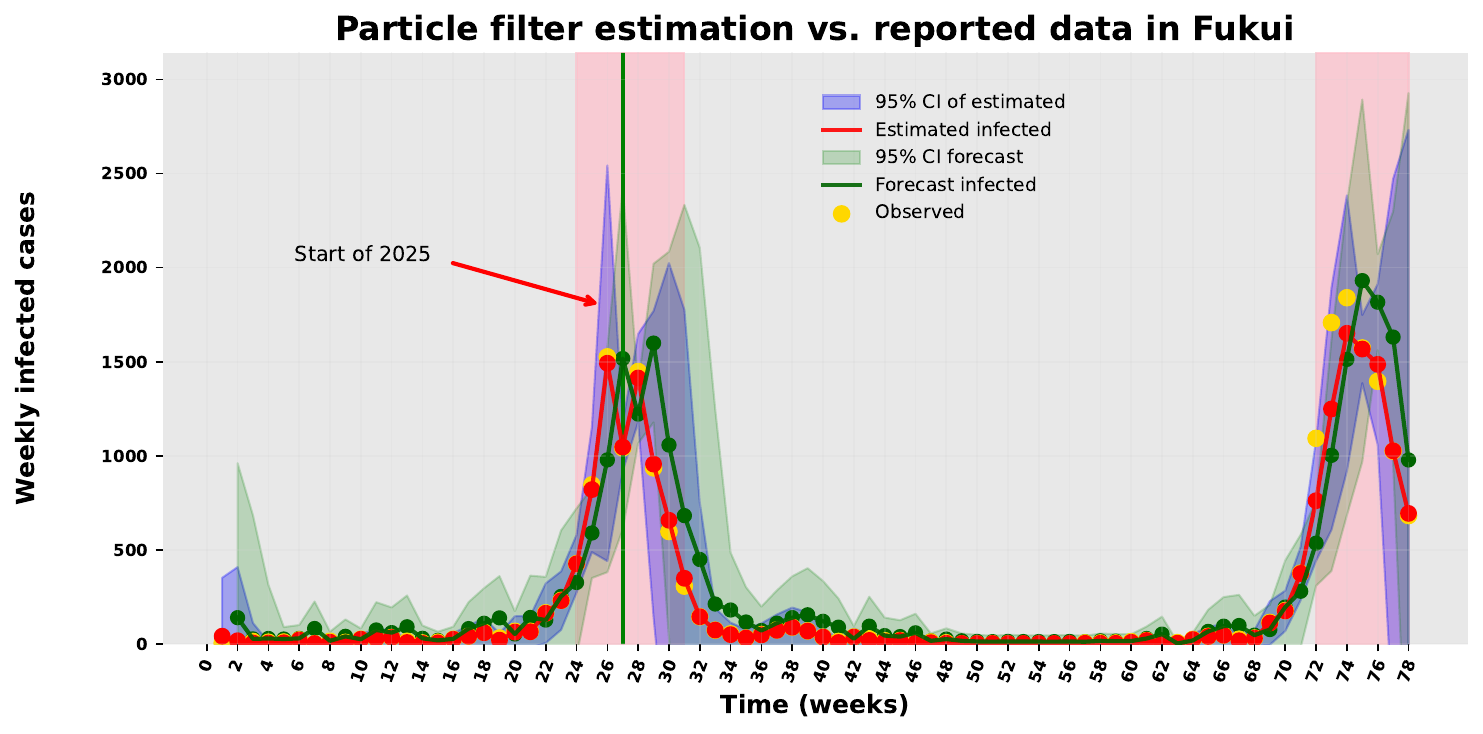}
%       \vspace{-1.0cm}
  \caption{One week ahead model forecast of weekly influenza cases in Fukui.}
    \label{forcast_fukui}
\end{figure}

The one-week ahead also forecasts for Kyoto and Fukui are shown in Figures~\ref{forcast_kyoto} 
and~\ref{forcast_fukui}, respectively. The epidemic peak in Fukui is captured well because the number 
of reported cases is lower than in Kyoto. 
In contrast, Kyoto shows a larger increase in cases, which makes the peak harder to capture accurately.

\begin{figure}[htbp]
    \centering
    \begin{minipage}[b]{0.495\textwidth}
        \centering
        \includegraphics[width=\textwidth]{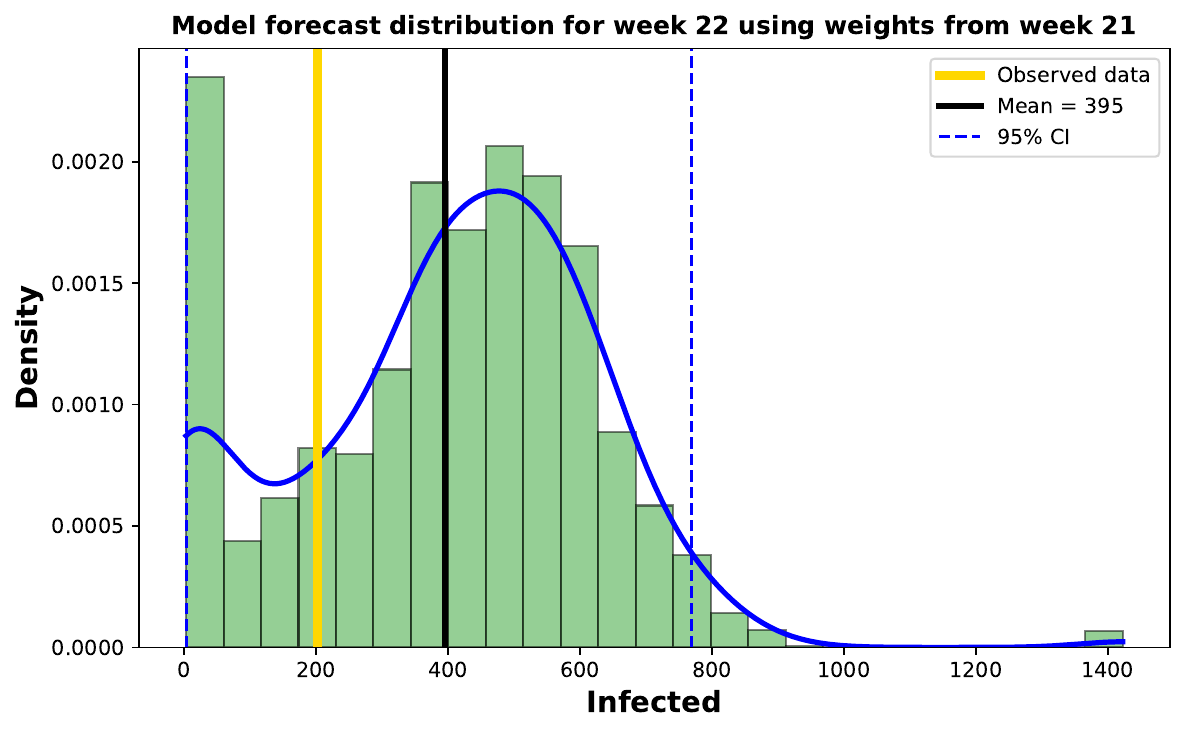}
    \end{minipage}
    \hfill
    \begin{minipage}[b]{0.495\textwidth}
        \centering
        \includegraphics[width=\textwidth]{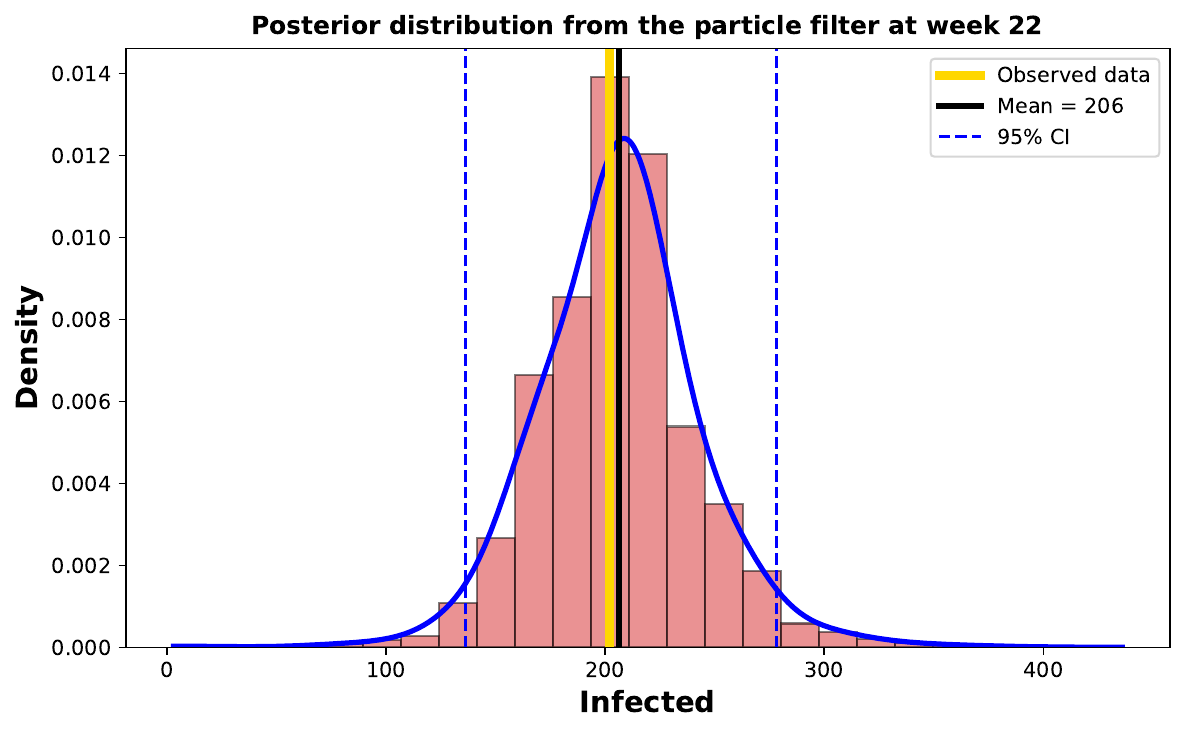}
    \end{minipage}
    \caption{One week ahead model forecast distribution and posterior distribution of particle filter of weekly influenza cases in Okinawa.}
    \label{posterior_forcats}
\end{figure}

Figure~\ref{posterior_forcats} shows the one week ahead model forecast distribution 
and the particle filter posterior distribution of weekly influenza cases in Okinawa at week $22$. 
The forecast distribution, generated using the weights from week $21$, is relatively broad, 
reflecting the uncertainty in predicting the next observation. 
After assimilating the reported data at week $22$, the posterior distribution becomes narrower and 
more concentrated around the observed incidence. 
This indicates that the particle filter update refines the forecast and reduces the uncertainty in the estimate.

\subsection{Transmission rate for the graph based model}

\begin{figure}[htbp]
    \centering
    \includegraphics[width=0.9\textwidth]{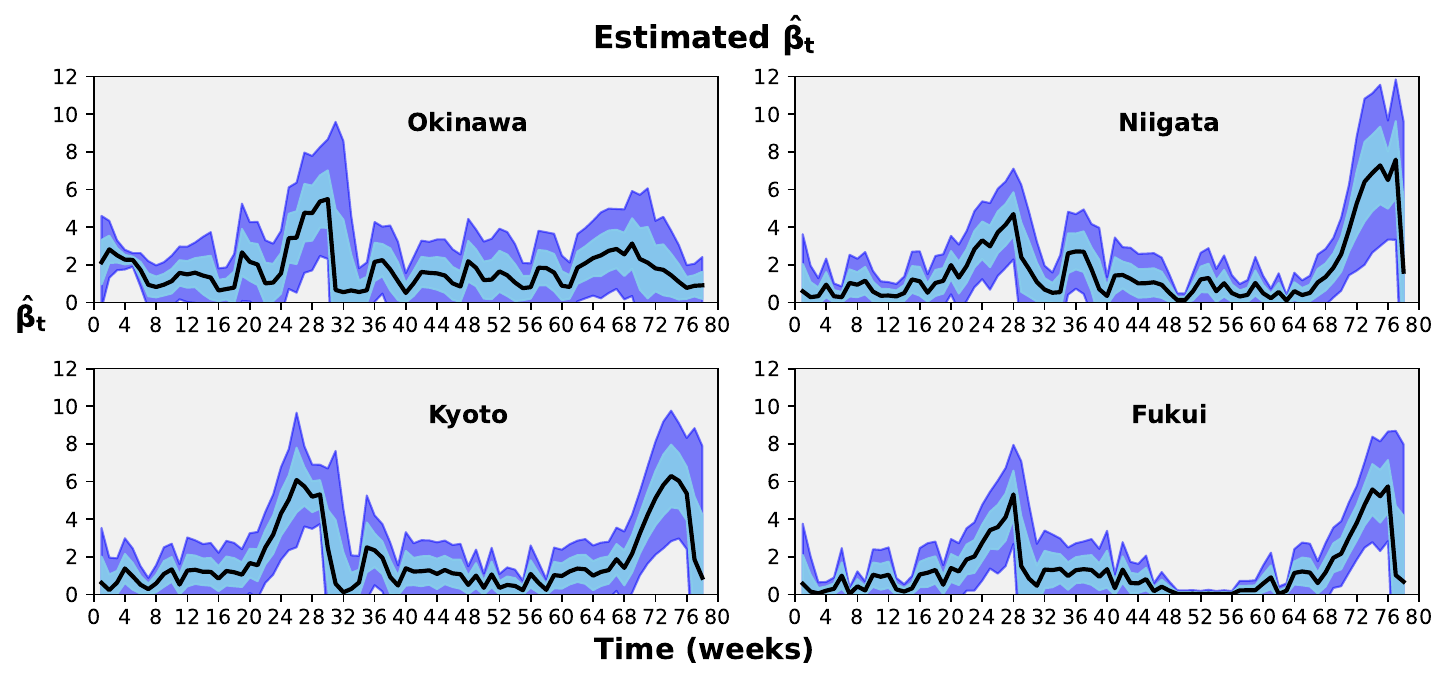}
%       \vspace{-1.0cm}
   \caption{Estimated transmission rate $\hat{\beta}_{t}$ for the graph-based stochastic model in four prefectures 
   of Japan from July 2024 to December 2025: Okinawa, Niigata, Kyoto, and Fukui. 
   The solid curve show the estimated transmission rate values. The shaded regions show the confidence intervals, 
   with blue indicating the $95\%$ confidence interval and sky blue indicating the $65\%$ confidence interval.}
    \label{estimated_betas}
\end{figure}

In this subsection, we present the estimated transmission rate $\hat{\beta_{t}}$ for these four prefectures in Japan 
for the same period. The methodology is presented in Section \ref{PF_method1_saection}. 
Figure~\ref{estimated_betas} presents the estimated time-varying transmission rate $\hat{\beta}_{t}$ 
for Okinawa, Niigata, Kyoto, and Fukui. 
In all regions, higher uncertainty is observed during the peak stages of the epidemic, with sudden jumps in the data. 
As the outbreak progresses and approaches the epidemic peaks, the PF effectively captures the changes in transmission intensity. 
It also follows the subsequent decrease, although with a slight delay. 
The estimated transmission rates clearly reflect the timing of the epidemic waves observed in each region, 
with significant increases during major outbreak periods and near-zero values during low-transmission phases. 
Therefore, the results demonstrate the ability of the PF to track temporal variations in transmission dynamics 
across different prefectures.

\begin{figure}[htbp]
\centering
\begin{overpic}[width=0.30\textwidth,height=0.15\textheight]{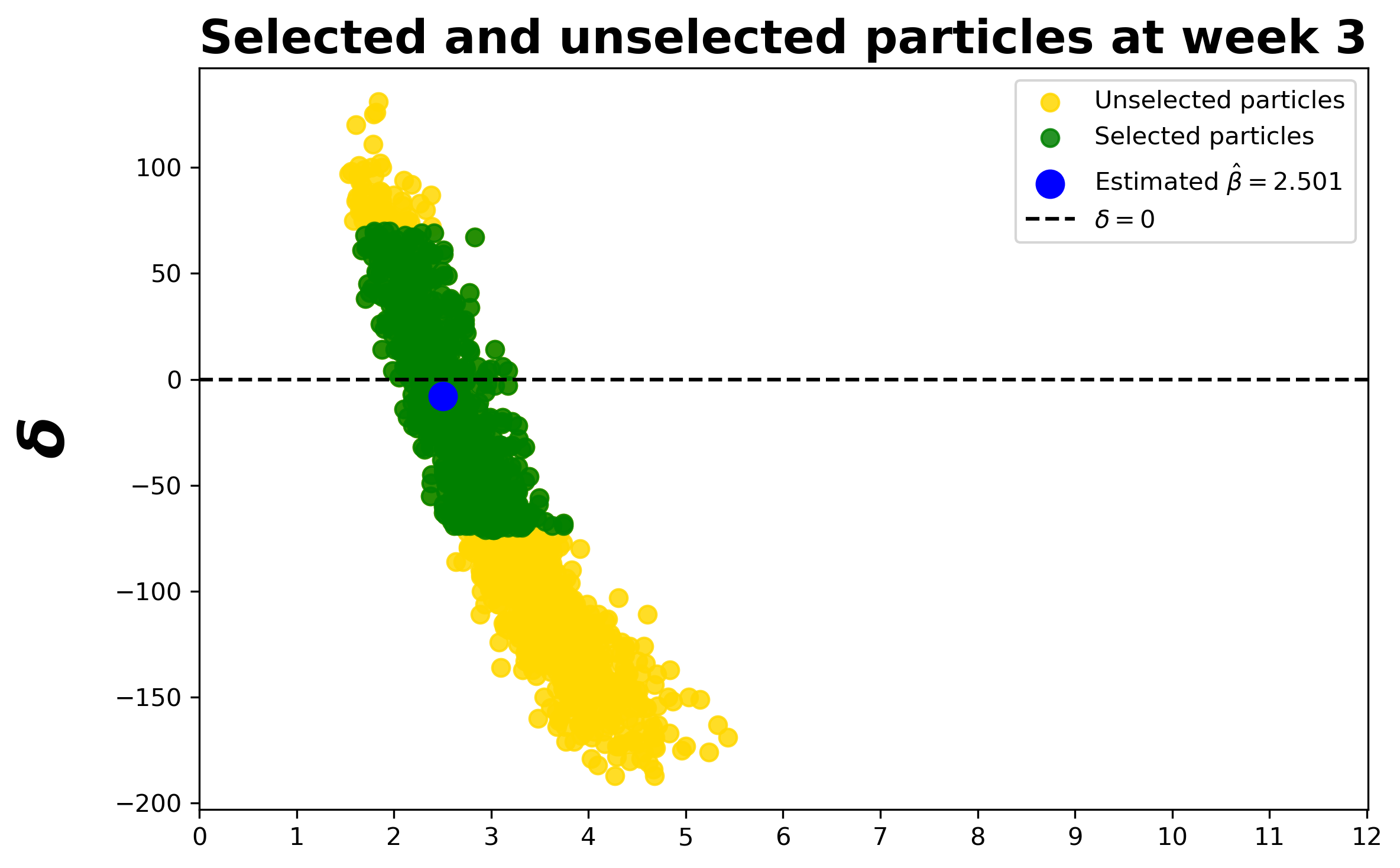}
\end{overpic}
\begin{overpic}[width=0.30\textwidth,height=0.15\textheight]{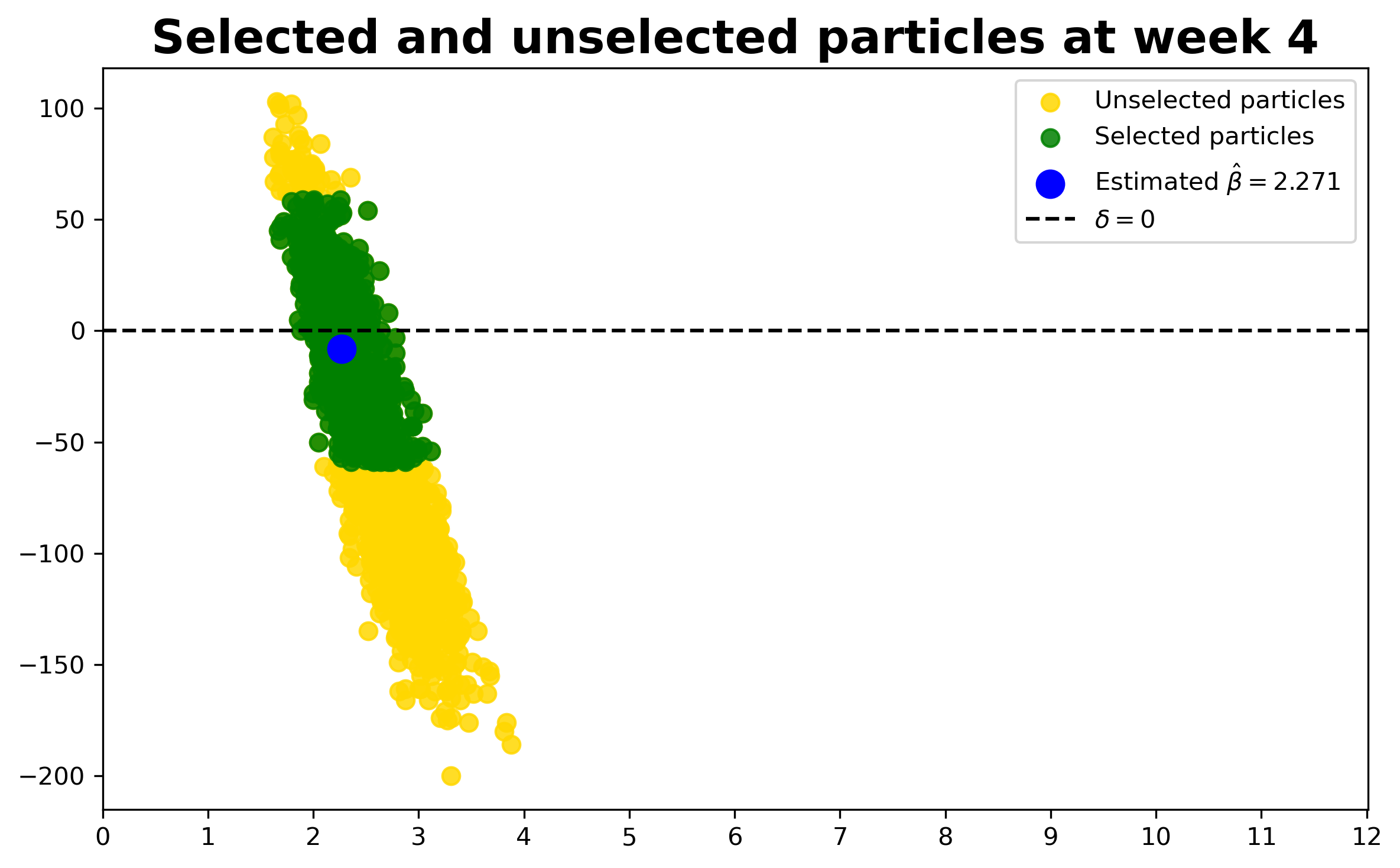}
\end{overpic}
\begin{overpic}[width=0.30\textwidth,height=0.15\textheight]{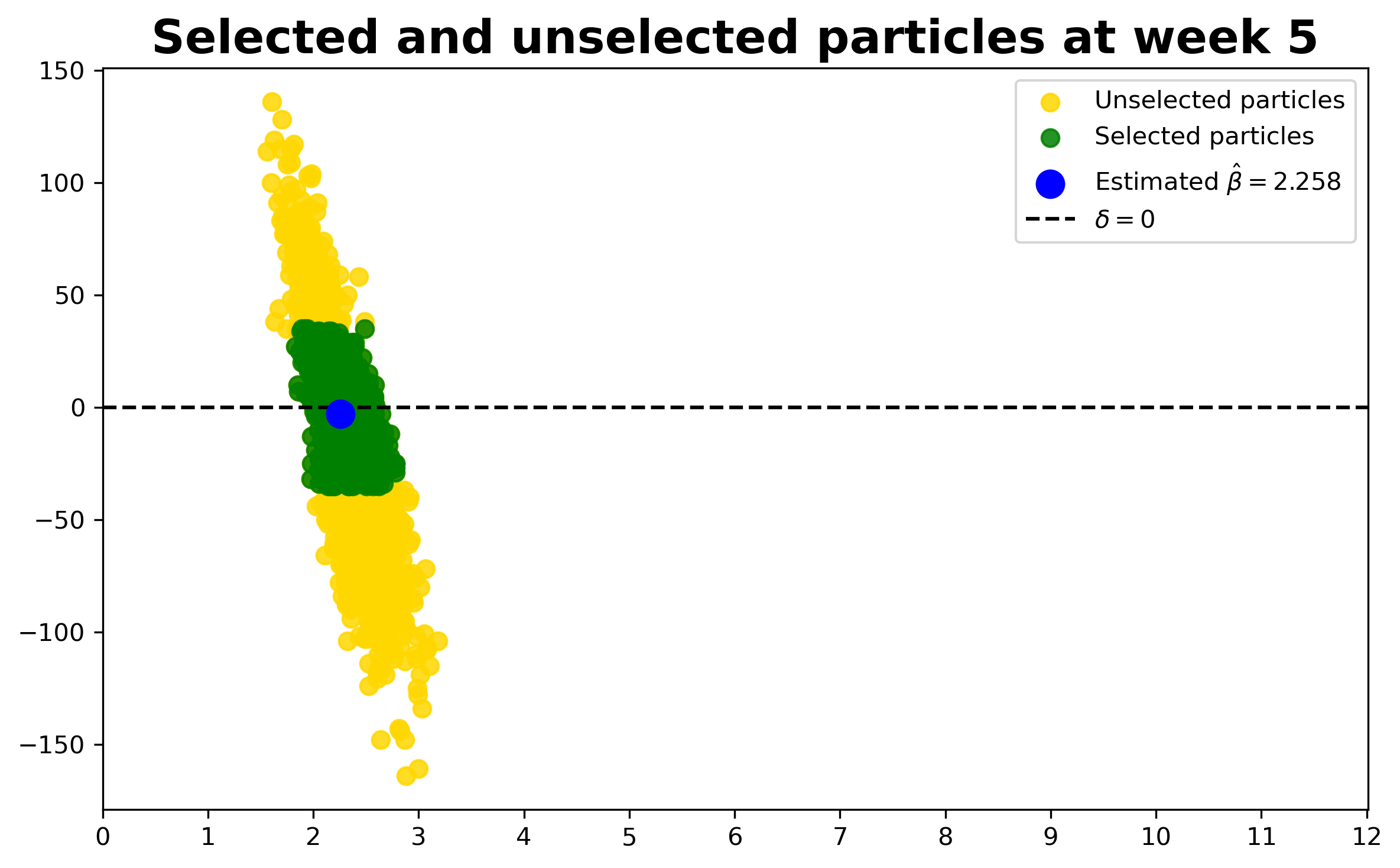}
\end{overpic}

\vspace{0.1cm}

\begin{overpic}[width=0.30\textwidth,height=0.15\textheight]{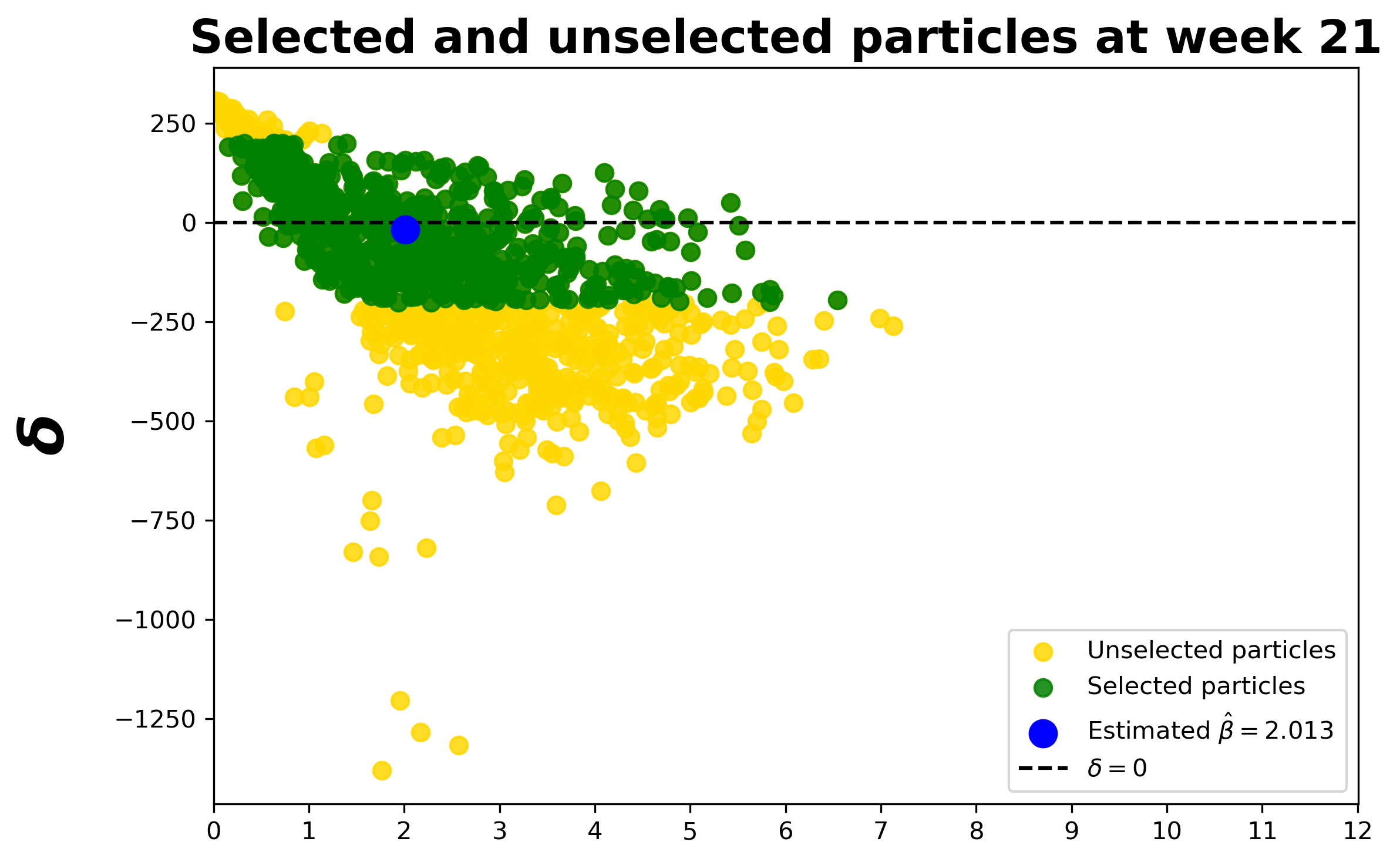}
\end{overpic}
\begin{overpic}[width=0.30\textwidth,height=0.15\textheight]{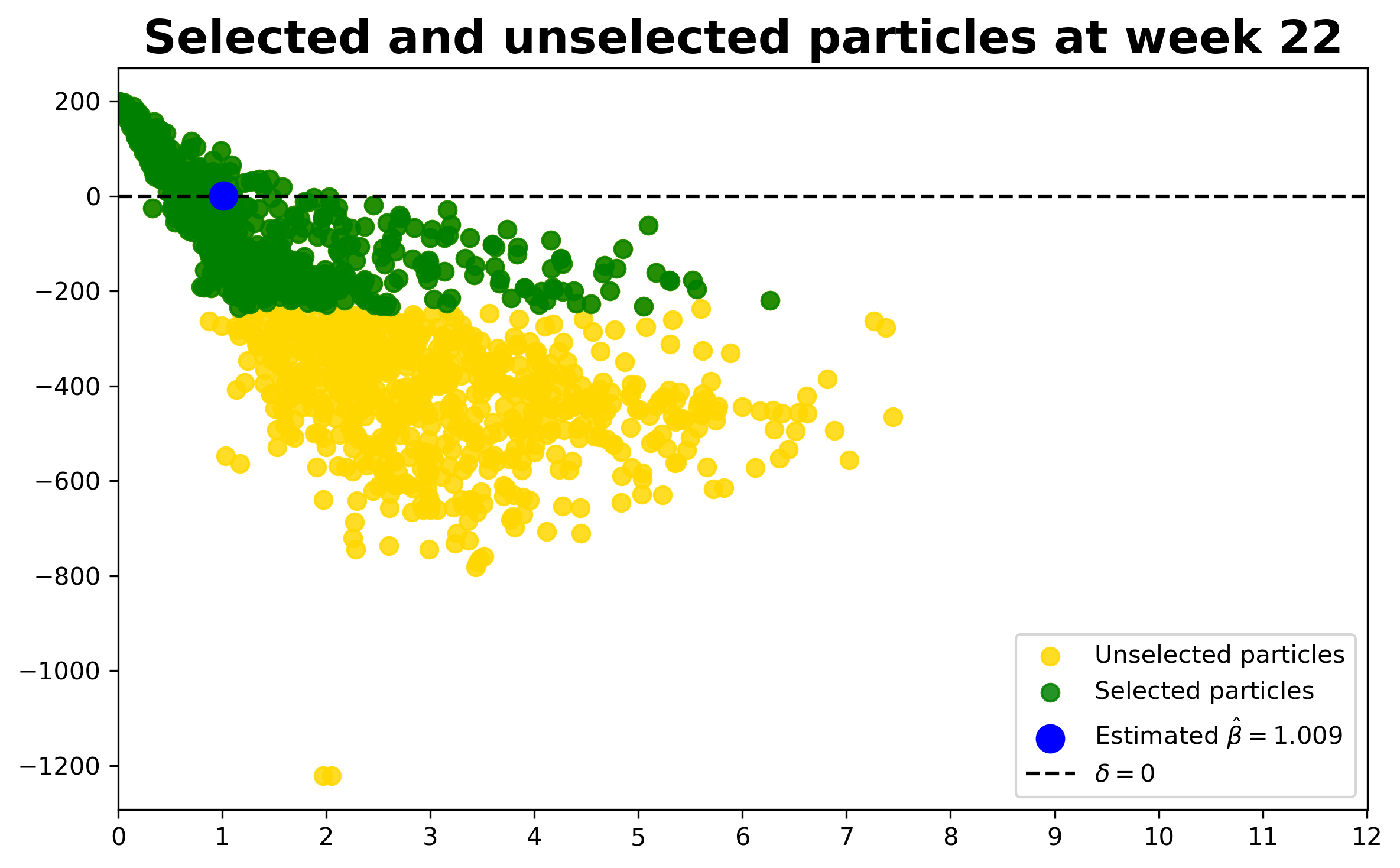}
\end{overpic}
\begin{overpic}[width=0.30\textwidth,height=0.15\textheight]{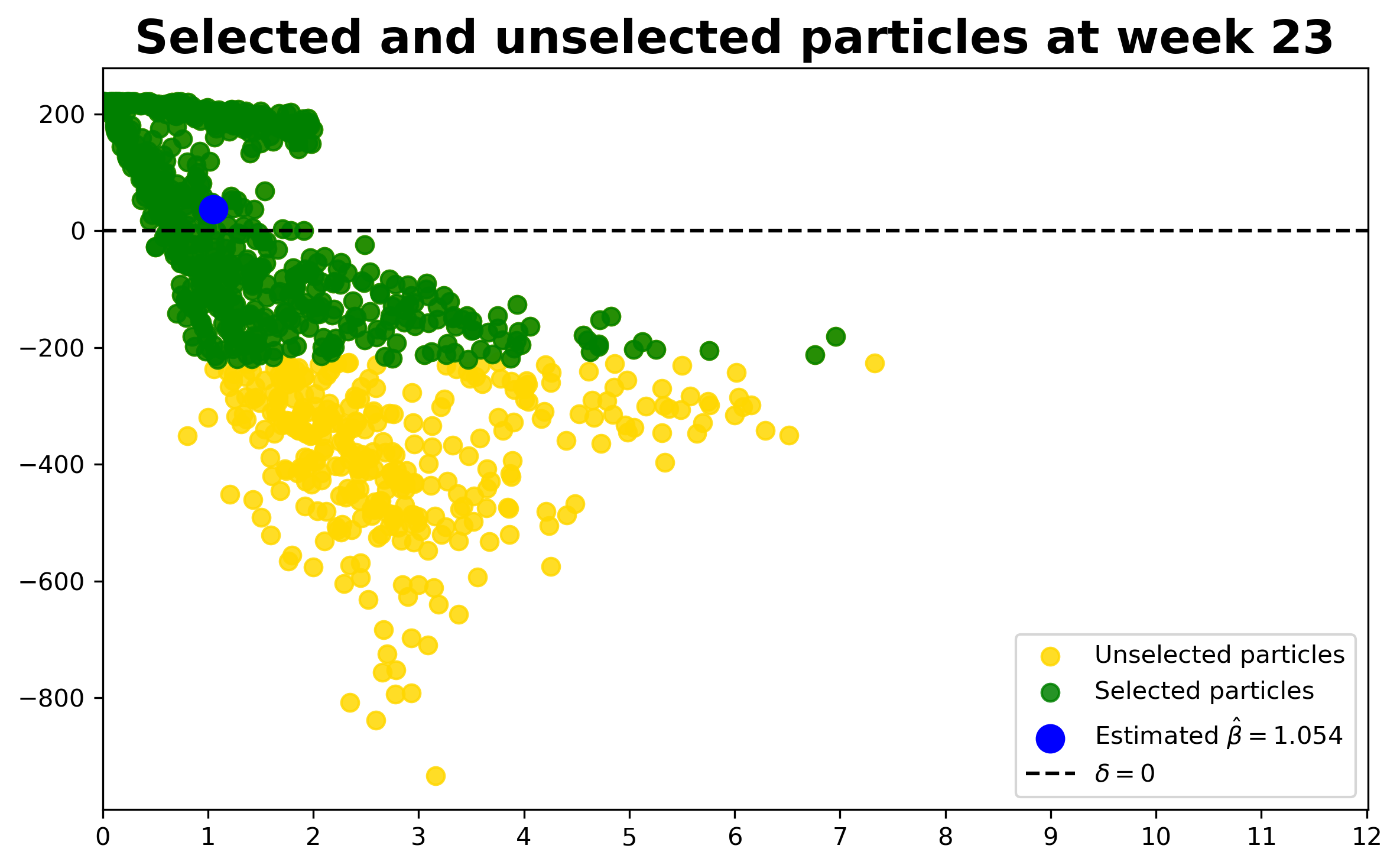}
\end{overpic}

\vspace{0.1cm}

\begin{overpic}[width=0.30\textwidth,height=0.15\textheight]{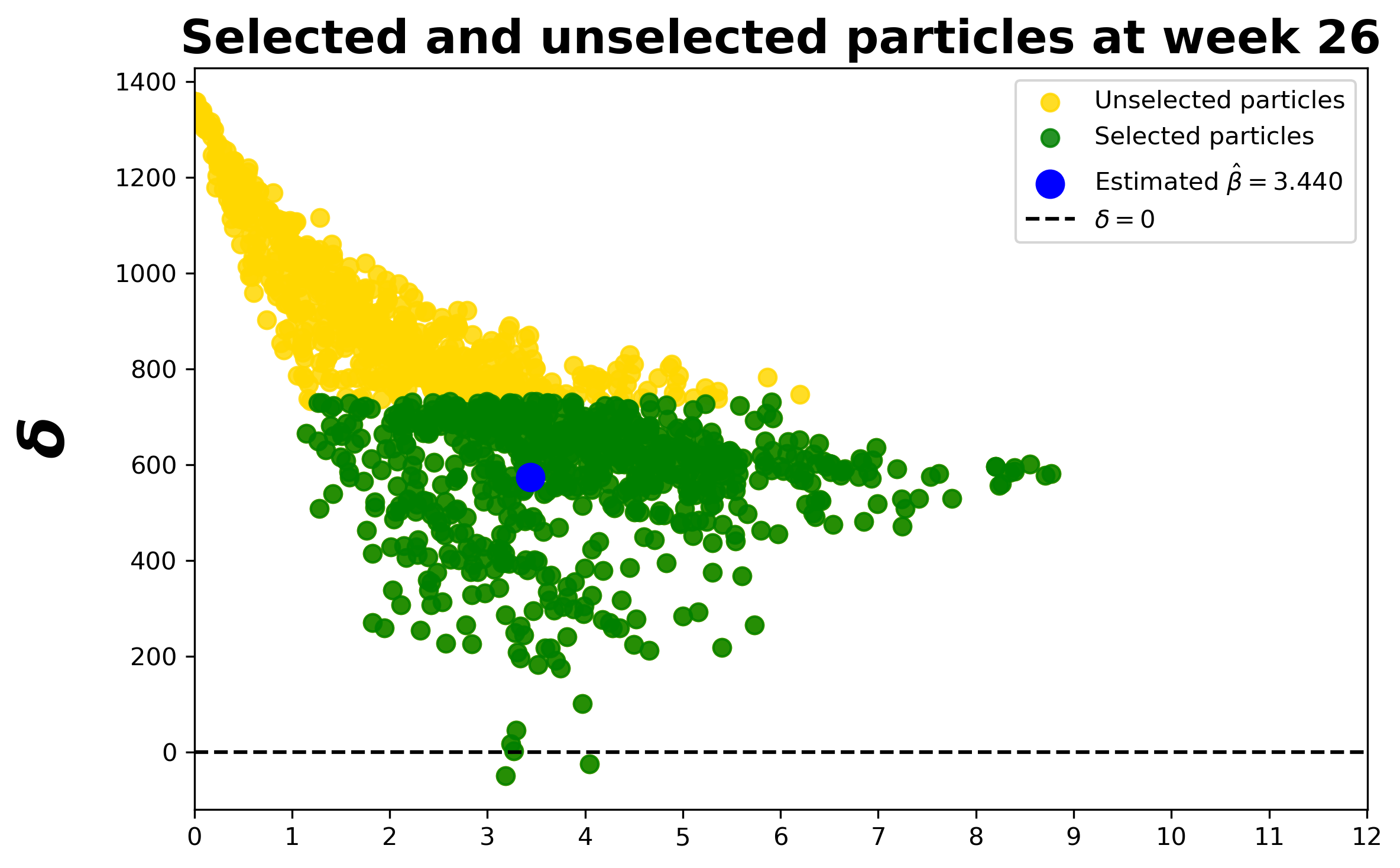}
\end{overpic}
\begin{overpic}[width=0.30\textwidth,height=0.15\textheight]{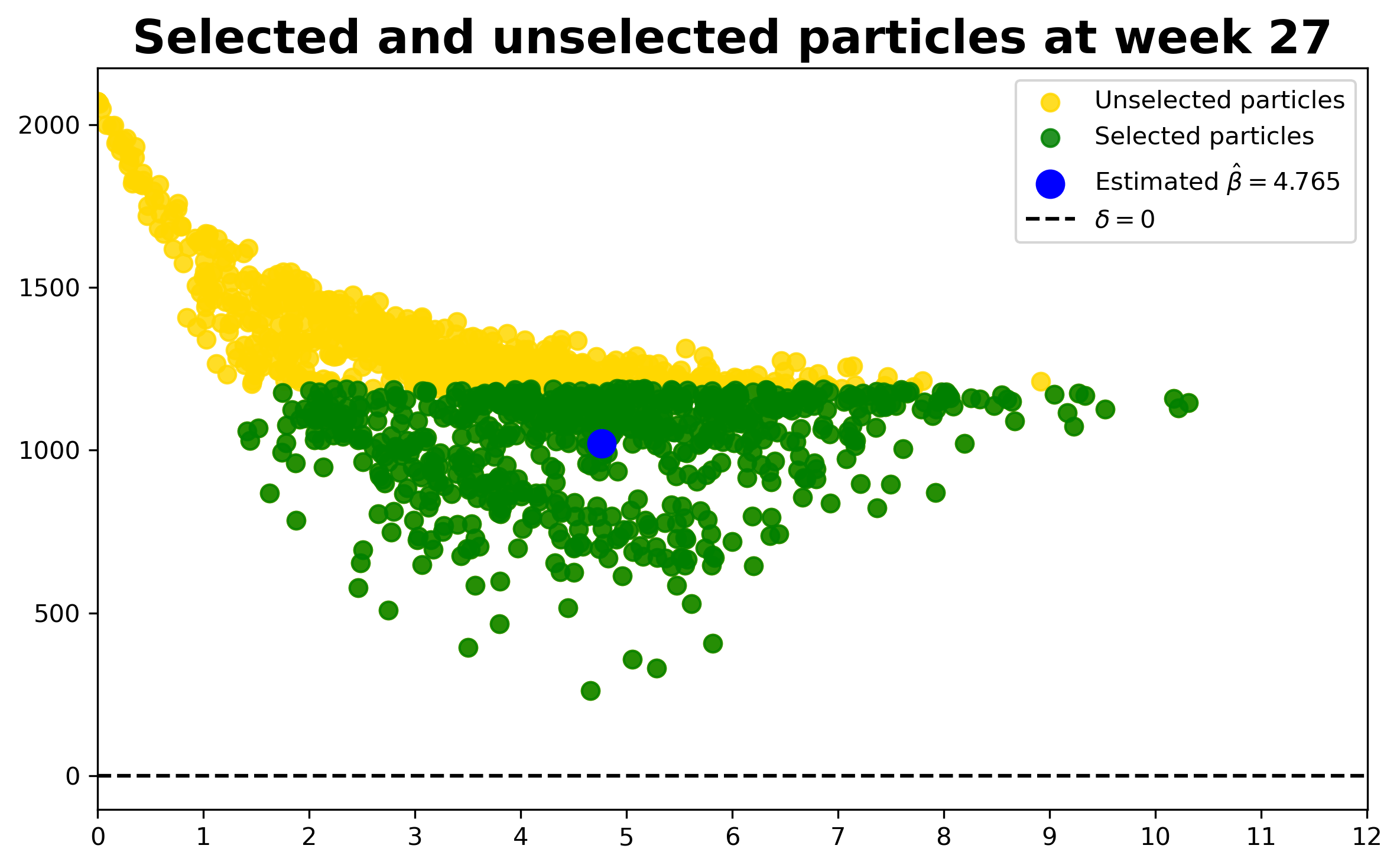}
\end{overpic}
\begin{overpic}[width=0.30\textwidth,height=0.15\textheight]{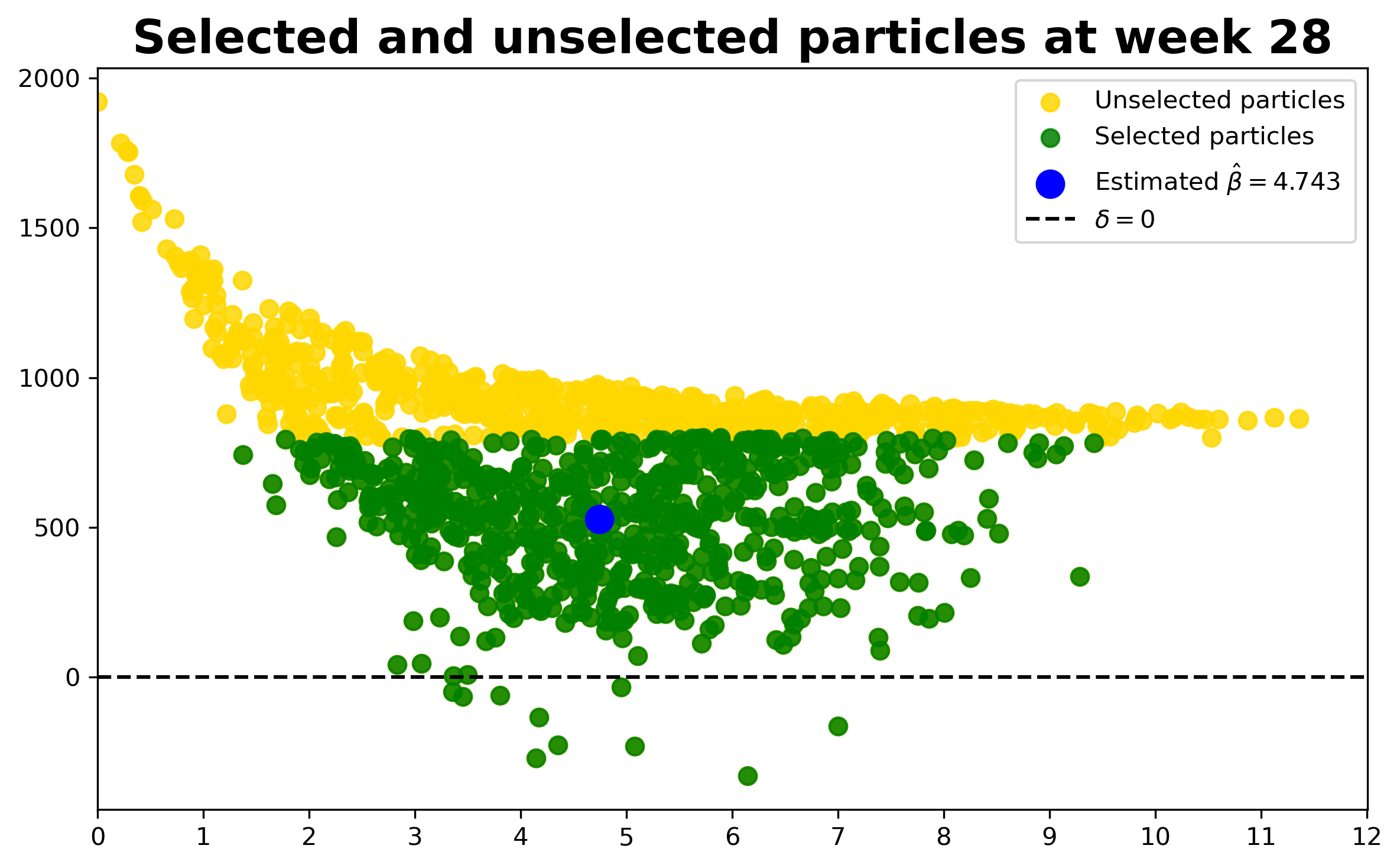}
\end{overpic}

\vspace{0.1cm}

\begin{overpic}[width=0.30\textwidth,height=0.15\textheight]{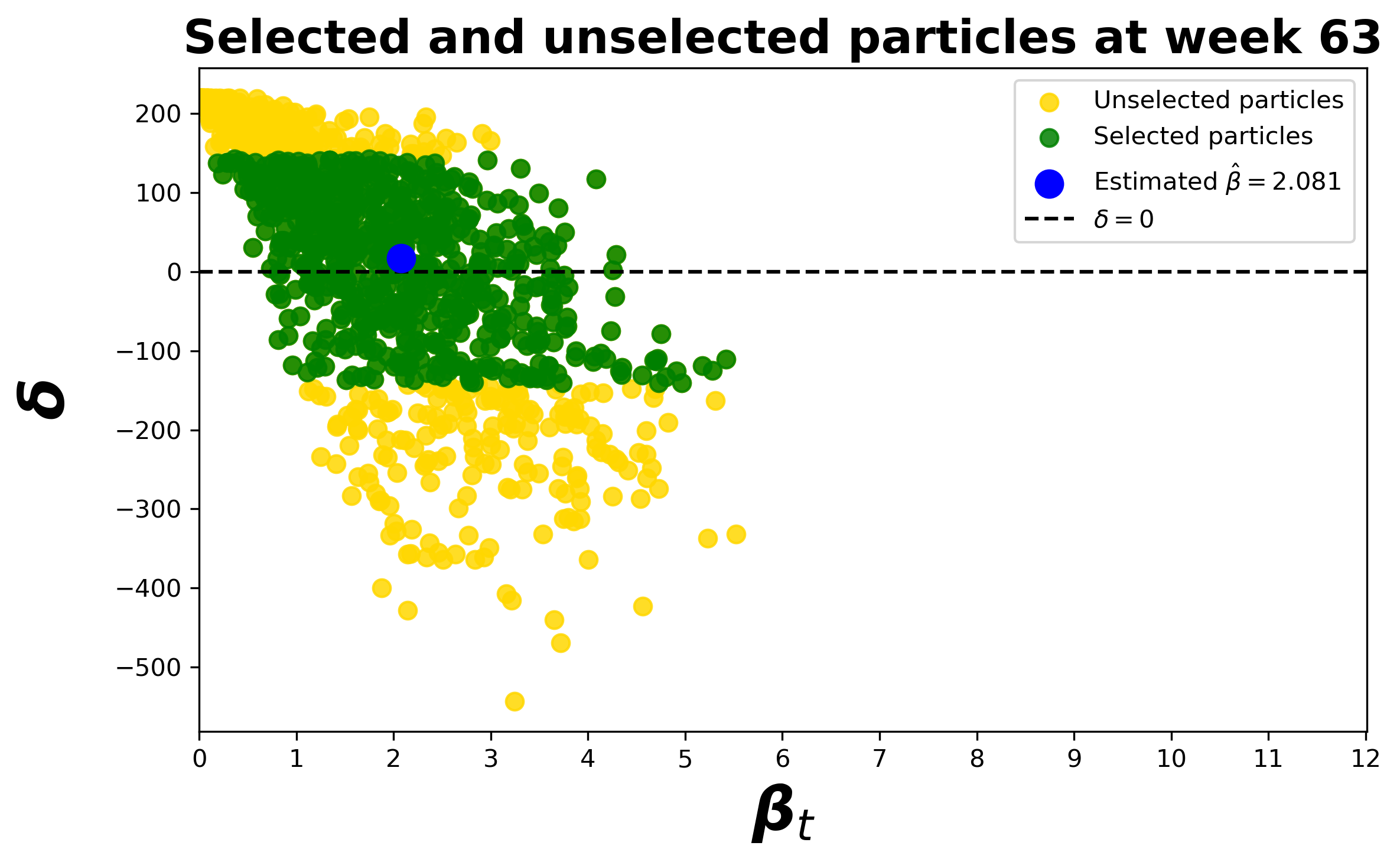}
\end{overpic}
\begin{overpic}[width=0.30\textwidth,height=0.15\textheight]{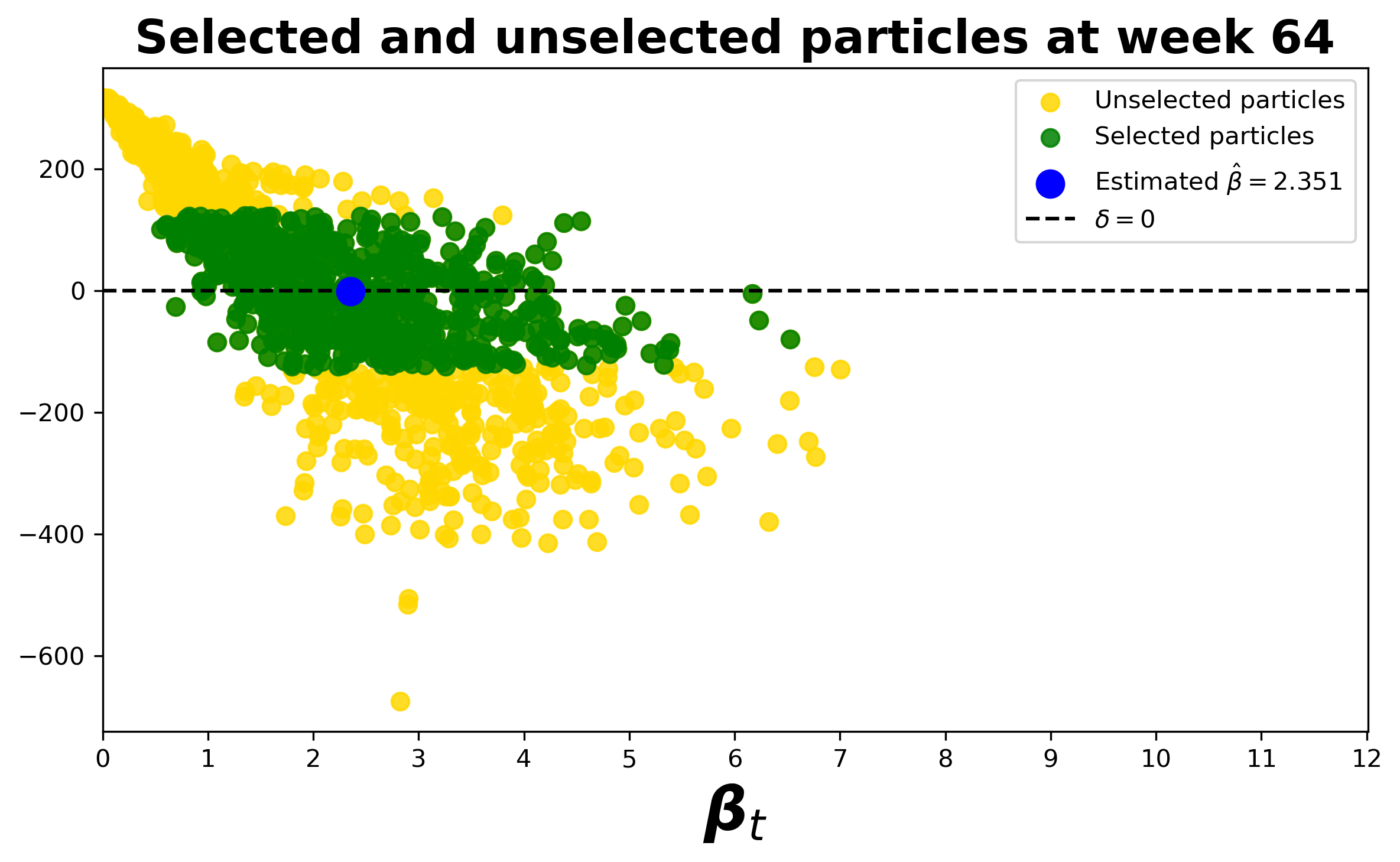}
\end{overpic}
\begin{overpic}[width=0.30\textwidth,height=0.15\textheight]{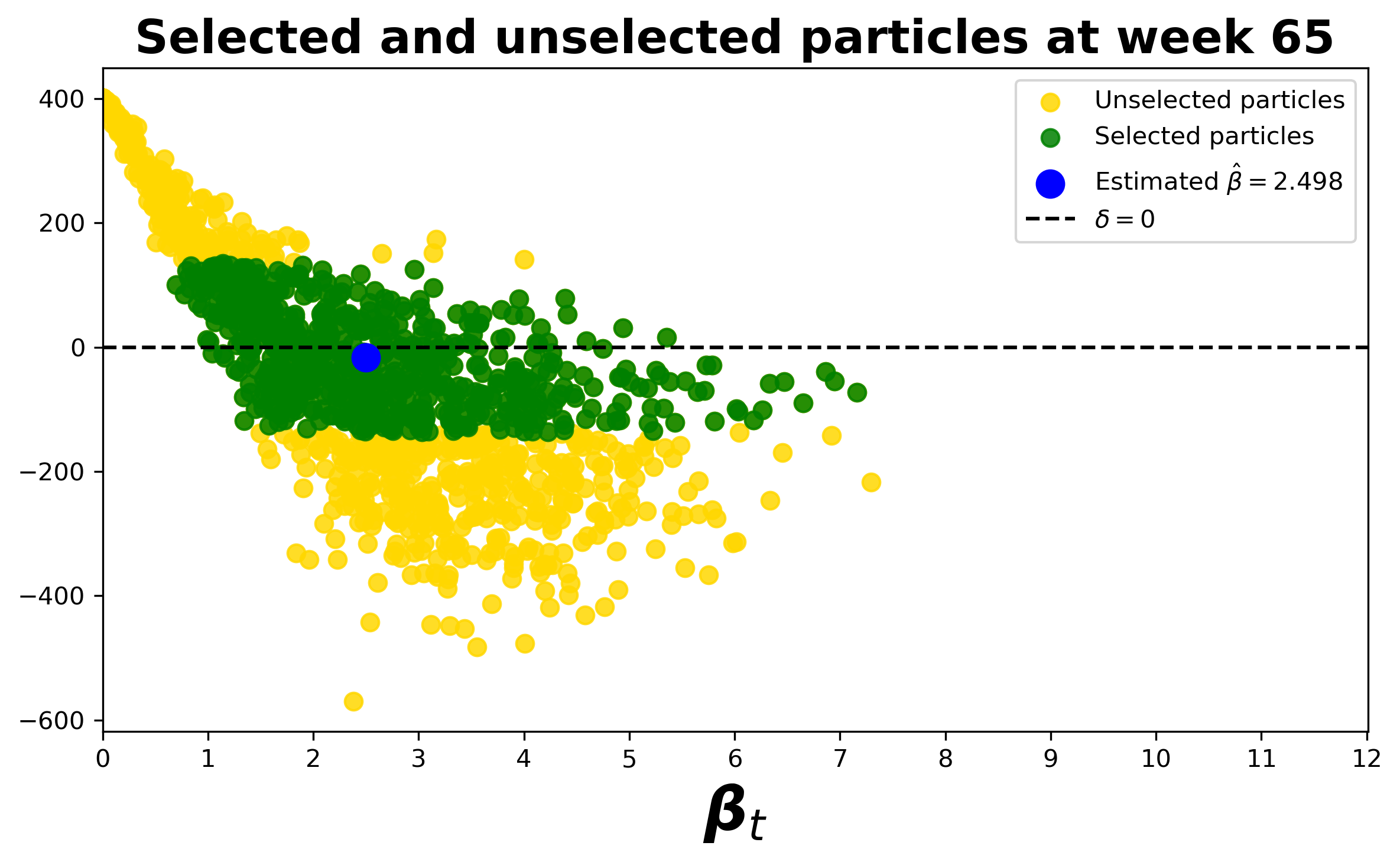}
\end{overpic}
\caption{Selected and unselected particles at different weeks using the absolute distance criterion. 
Green represents selected particles, yellow represents unselected particles, 
the blue point shows the estimated $\hat{\beta}_{t}$, and the dashed line denotes $\delta=0$.}
\label{particle_distance}
\end{figure}

In order to illustrate the distribution of $\beta_t$ and the selection process of the particles in our model, 
we provide in Figure \ref{particle_distance} a comparison between the observed data and the simulated data for Okinawa 
at different observation time. In this figure, each dot represents a particle by its coordinates 
$\big(\beta_{t_k}^{(i)}, \delta_{t_k}^{(i)}\big)$
with 
\begin{equation*}
\delta^{(i)}_{t_k} =  y_{t_k} - x_{t_k}^{(i)}, \qquad \quad \forall i \in \{1, \dots, N_{p}\}.
\end{equation*} 
The green particles are selected according to the formula defined in Method 1 in \eqref{absolute_distance}, 
while the unselected particles are represented in yellow. These selected particles are used to estimate 
the state and parameter at the current time. Based on their fitness, we also decide the sampling for the next step.
In the figure, the blue point shows the estimated $\hat{\beta}_{t}$, and the dashed line denotes $\delta=0$.

At week 4, as shown in Figure~\ref{particle_distance}, a sufficient number of particles satisfy the 
fitness condition given in equation \eqref{particles_fitness}. 
Therefore, in the transition from week 4 to week 5, step $(v)$ is applied, which uses a small perturbation 
around the selected particles. This indicates that the selected particles already provide a good representation 
of the state and parameter, so only a local update is needed. In contrast, at week 26 and 27, the particles 
do not satisfy the required fitness condition. In this case, step $(v)$ is not applied. Instead, step $(v')$ 
is used for the transition from week 26 to week 27 and then to week 28. 
This larger perturbation allows a wider exploration of the parameter 
space when the current selected particles are not good enough.

\subsection{Transmission rate for deterministic model}

\begin{figure}[htbp]
    \centering
    \begin{subfigure}[b]{0.45\textwidth}
        \centering
        \includegraphics[width=\textwidth]{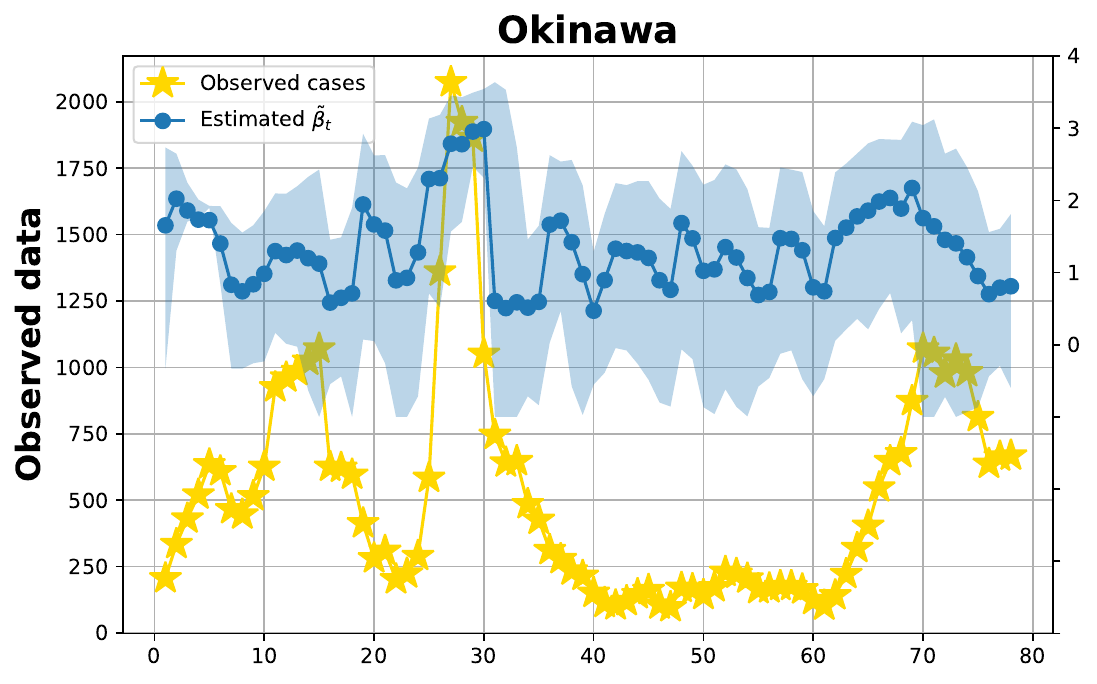} 
    \end{subfigure}
    \begin{subfigure}[b]{0.45\textwidth}
        \centering
        \includegraphics[width=\textwidth]{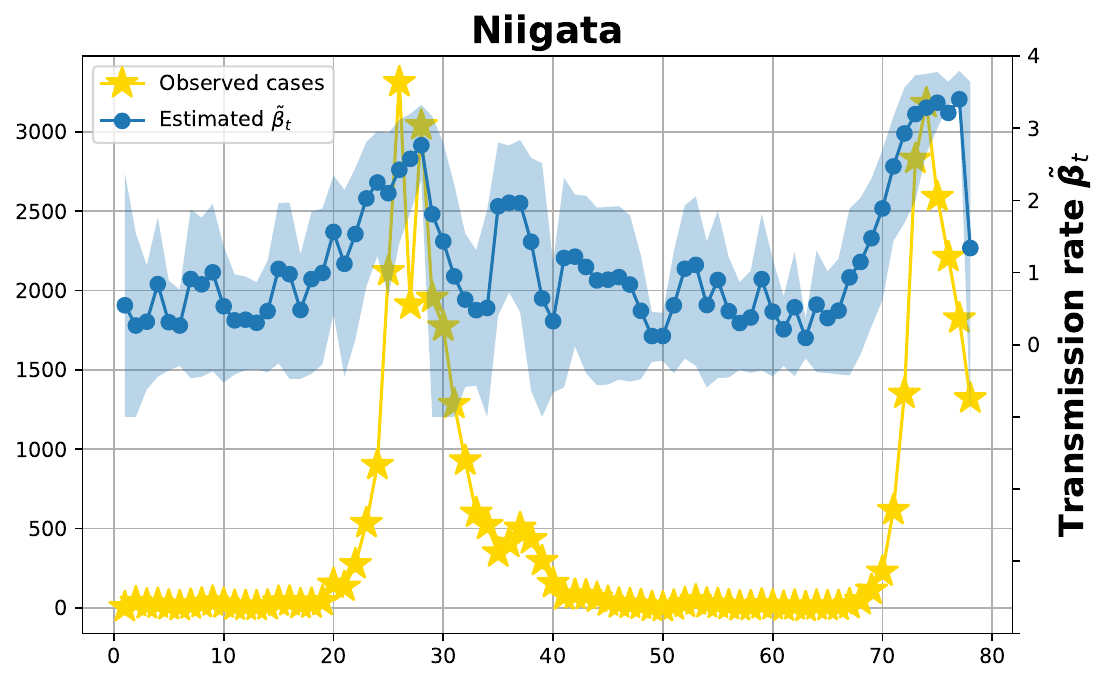} 
    \end{subfigure}
    \vspace{0.2cm}
    \begin{subfigure}[b]{0.45\textwidth}
        \centering
        \includegraphics[width=\textwidth]{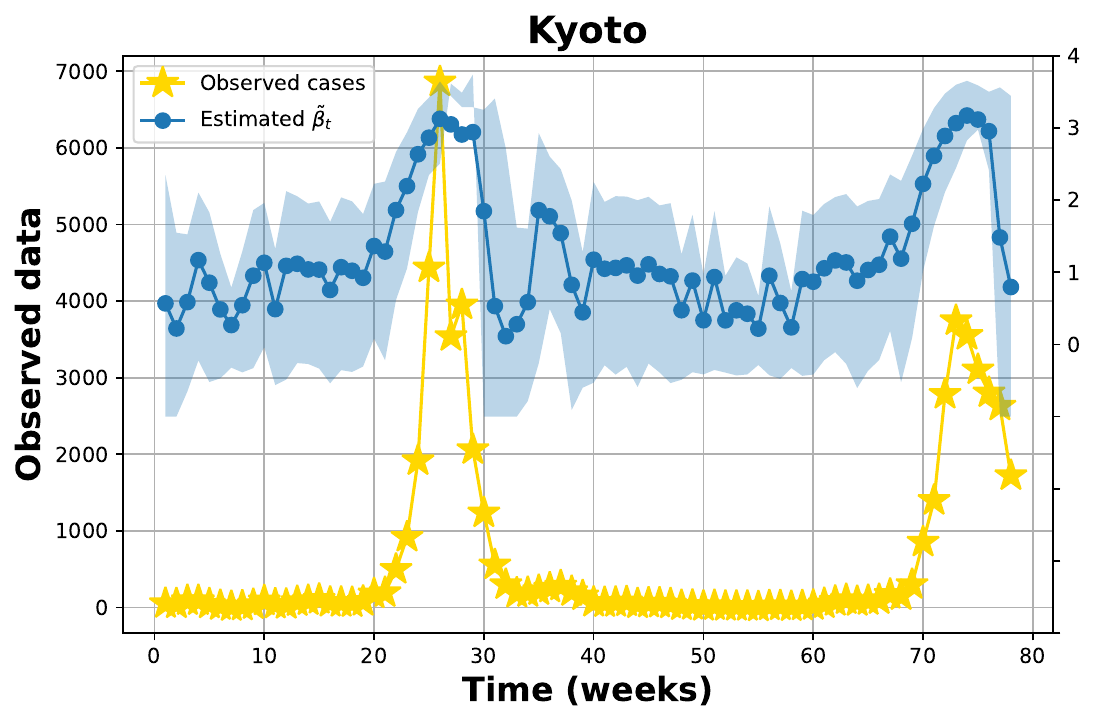} 
    \end{subfigure}
    \begin{subfigure}[b]{0.45\textwidth}
        \centering
        \includegraphics[width=\textwidth]{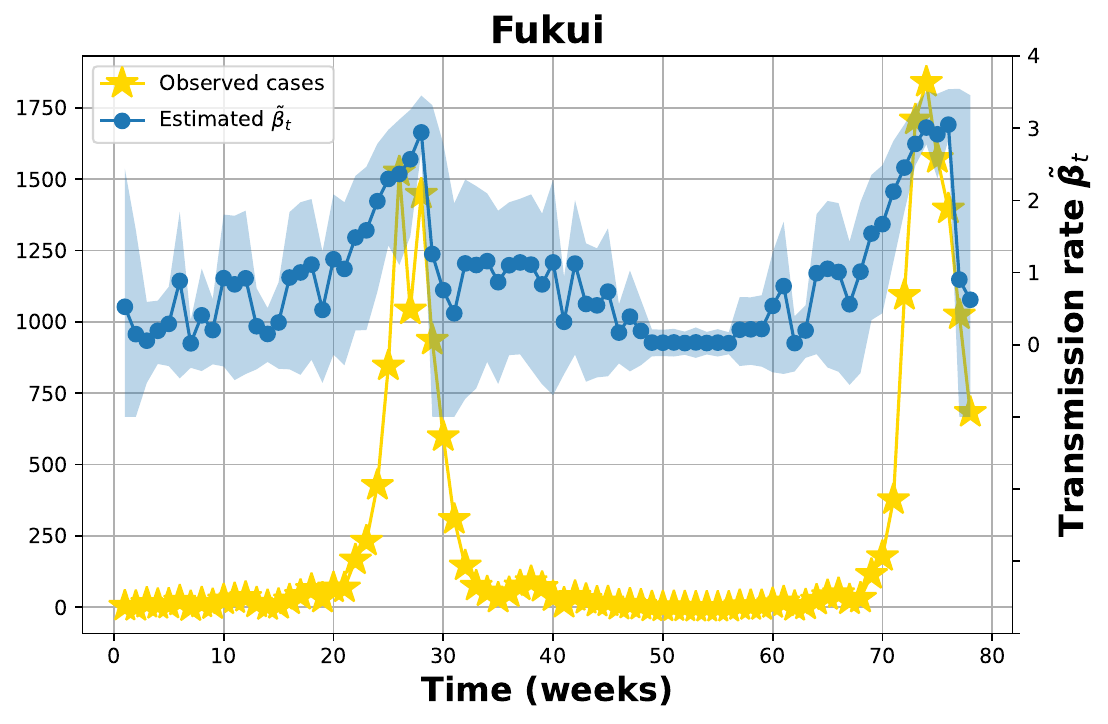} 
    \end{subfigure}
    \caption{Estimated transmission rate $\hat{\beta}_{t}$ for the deterministic model for four prefectures 
    in Japan from July 2024 to December 2025. The blue curves show the estimated transmission rate values. 
    The shaded regions illustrate the confidence intervals, with blue indicating the $95\%$ interval confidence interval.
    For comparison, the observed infected cases are also represented in these figures.}
    \label{transmission_rates} 
\end{figure}

We finally examine a method for converting the graph-based transmission rate into the SIR model-based transmission rate. 
The transmission rate $\tilde{\beta}_{t}$ in the deterministic SIR model is commonly defined as the product of the contact rate 
and the probability of transmission per contact, i.e., $\tilde{\beta}_{t} = C \times p$, see \cite{Roy2023} for further details.
From Section~\ref{model_section}, we know that the probability of a susceptible individual becoming infected is
\begin{equation}
P(S \to I)=1 - \exp\!\left(-\hat{\beta_{t}}\frac{\I^{\kappa}_{t}}{4}\right).
\label{probability}
\end{equation}
From Figure \ref{model_simulation}, we observe that most susceptible individuals have one infected neighbor. 
Therefore, we set $\I^{\kappa}_{t} = 1$ and take the possible contact rate to be $C = 4.0$. 
\begin{equation}
\tilde{\beta}_{t}= 4 (1 - e^{-\frac{\hat{\beta_{t}}}{4}}).
\label{tran_rate_det}
\end{equation}
The resulting transmission rate $\tilde{\beta}_{t}$ is provided for the four prefectures in 
Figure \ref{transmission_rates}. For comparison, the observed data are also shown in the same figures.

\section{Conclusion}\label{conclusion}

Classical compartmental epidemic models, such as the SIR ODE model, often assume homogeneous mixing of the population. 
This means that every individual has an equal probability of interacting with every other individuals. 
However, this assumption is unrealistic because individuals typically interact within limited social networks. 
Graph-based or agent-based models address this limitation by explicitly representing individuals as nodes 
and their interactions as edges. This structure allows the model to capture heterogeneous contact patterns, 
spatial interactions, and individual behavior. 
In this study, we develop a graph based model for infectious disease transmission. 
Our analysis focuses on a two-dimensional lattice with an infinite number of nodes 
and the transmission occurs only between a node and its nearest neighbors. 
Since the model is defined on a 2D lattice, it allows a simple visualization of the three epidemiological states in the system: 
susceptible, infected, and recovered. The model shown in Figure \ref{model_simulation}, 
with four nearest-neighbor connections, simulates disease propagation on a discrete domain of nodes 
representing individuals in contact with one another.

In these graph-based epidemic models, such as two-dimensional lattice or network models, 
errors in the initial states and model parameters can lead to large uncertainties in the predicted disease dynamics. 
Because these models often contain various uncertain parameters related to contact interactions between individuals, 
making calibration with real epidemic data remains challenging. 
Data assimilation provides a framework for integrating observational data with model simulations 
in order to estimate hidden states and uncertain parameters and improve the accuracy of epidemic forecasts. 
We presented two particle filter algorithms to address the problem of state and parameter estimation 
in biological phenomena modeled by network-based systems. 
The use of PFs provided estimates of the state and parameters, as well as for the associated  uncertainties. 
The proposed PF-based data assimilation framework maximizes the likelihood of hidden states and parameters 
of the model.

Our approach differs from standard PF by integrating a mechanism that approximates 
the posterior distribution using a fixed window size of recent observations and different resampling approaches. 
It reduces computational demands while maintaining parameter and state estimation accuracy. 
The standard PF algorithm has the issue of loss of particle diversity caused by particle degeneracy and resampling. 
This makes it difficult for particle samples to accurately represent the true probability density function of the system state. 
Our algorithm improves this by replicating the best particles and removing the worst ones, 
to accelerating convergence while maintaining particle effectiveness. 
This approach improves the ability of the algorithm to estimate unknown states with improved speed and accuracy. 
This adaptation  makes PF particularly suitable for dynamic disease modeling, where timely responses are critical. 

We demonstrated the effectiveness of PF algorithms and validated their performance on both synthetic 
simulations and real-world Influenza data from Japan from July 2024 to December 2025.
Numerical experiments indicate that window size $\tau$ significantly affects the accuracy of estimation. 
In general,reducing the window size $\tau$ allows the posterior distribution to be updated more frequently, 
which improves estimation accuracy but increases the computational cost. 
We also performed one week ahead forecasting simulations based on the current week data.
The forecasts were generally accurate during the early and decreasing stages.

We finally acknowledge two limitations in this work. 
Firstly, we did not take into account agent mobility, meaning that all agents remain in fixed positions and do not change their locations. 
However, we don't expect this limitation to play a major role, since the initial position of our agent is not determined: 
by definition, the transmission of the disease will always take place between "interacting" agents, no mater where they are 
originally located in space. Once this interaction is taking place, these agents become neighbors with respect to the epidemic propagation,
and therefor they become neighbors in our framework.
The second limitation is about the maximum number of agents that one infected agent can infect. In our 2D model, this number is 
limited to $3$ (at least one neighbor is already infected). In order to increase this number, one would have to consider $n$-dimensional
model, with $n>2$, or add edges on the 2D lattice. Since this would have a negative impact on the easy visualization of our model,
we haven't investigated further in this direction.

\section*{\textbf{Credit authorship contribution statement}}

\textbf{I. Haq}: Writing – original draft, Software, Methodology, Conceptualization; \textbf{S. Richard}: Writing – editing, Methodology, Conceptualization, Formal analysis 

\section*{\textbf{Data availability}}

All the data used in this article are available at \textcolor{blue}{\url{https://id-info.jihs.go.jp/surveillance/idwr/provisional/2025/47/index.html}}.

\section*{\textbf{Code availability}}
The code supporting the findings of this study will be made publicly available upon acceptance, and the GitHub repository link will be added here.

\section*{\textbf{Funding}}

This research received no specific grant from any funding agency in the public, commercial, or not-for-profit sectors.

\section*{\textbf{Competing interests}}

We declare that we have no competing interests.

\end{document}